\documentclass[sigconf]{acmart}
\setcopyright{none} 
\renewcommand\footnotetextcopyrightpermission[1]{}
\acmConference[arxiv]{}{August}{2023}

\AtBeginDocument{%
  \providecommand\BibTeX{{%
    \normalfont B\kern-0.5em{\scshape i\kern-0.25em b}\kern-0.8em\TeX}}}

\usepackage{amsmath,amssymb,amsfonts}
\usepackage{algorithmic}
\usepackage{graphicx}
\usepackage{textcomp}
\usepackage{xcolor}
\usepackage[ruled,linesnumbered]{algorithm2e}
\usepackage{multirow}
\usepackage{color}
\usepackage{threeparttable}
\usepackage{subfigure}
\usepackage{tcolorbox}
\usepackage{url}
\usepackage{booktabs}
\newcommand{\tool}{\textsc{MalModel}\xspace}

\begin{document}

\title{MalModel: Hiding Malicious Payload in Mobile Deep Learning Models with Black-box Backdoor Attack}

\author{Jiayi Hua}
\affiliation{%
  \institution{The Hong Kong Polytechnic University}
  \state{Hong Kong}
  \country{China}
}
\email{jiayi.hua@connect.polyu.hk}

\author{Kailong Wang}
\authornote{Kailong Wang and Haoyu Wang are the corresponding authors. 
The postal address is: Huazhong University of Science and Technology, Luoyu Road 1037, Wuhan, Hubei, China.
Email: wangkl@hust.edu.cn, haoyuwang@hust.edu.cn.}
\affiliation{%
  \institution{Huazhong University of Science and Technology}
  \city{Wuhan}
  \country{China}
}
\email{wangkl@hust.edu.cn}

\author{Meizhen Wang}
\affiliation{%
  \institution{Huazhong University of Science and Technology}
  \city{Wuhan}
  \country{China}}
\email{mzwang@hust.edu.cn}

\author{Guangdong Bai}
\affiliation{%
  \institution{University of Queensland}
  \city{Brisbane}
  \country{Australia}
}
\email{g.bai@uq.edu.au}

\author{Xiapu Luo}
\affiliation{%
 \institution{The Hong Kong Polytechnic University}
 \city{Hong Kong}
 \country{China}}
 \email{csxluo@comp.polyu.edu.hk}

\author{Haoyu Wang}
\authornotemark[1]
\affiliation{%
  \institution{Huazhong University of Science and Technology}
  \city{Wuhan}
  \country{China}}
\email{haoyuwang@hust.edu.cn}


\begin{abstract}
 Mobile malware has become one of the most critical security threats in the era of ubiquitous mobile computing.   
  Despite the intensive efforts from security experts to counteract it, recent years have still witnessed a rapid growth of identified malware samples. 
  This could be partly attributed to the newly-emerged technologies that may constantly open up under-studied attack surfaces for the adversaries. 
  One typical example is the recently-developed mobile machine learning (ML) framework that enables storing and running deep learning (DL) models on mobile devices. Despite obvious advantages, this new feature also inadvertently introduces potential vulnerabilities (e.g., on-device models may be modified for malicious purposes).

 In this work, we propose a method to generate or transform mobile malware by hiding the malicious payloads inside the parameters of deep learning models, based on a strategy that considers four factors (layer type, layer number, layer coverage and the number of bytes to replace). 
 Utilizing the proposed method, we can run malware in DL mobile applications covertly with little impact on the model performance~(i.e., as little as 0.4\% drop in accuracy and at most 39ms latency overhead). We test our method on seven different models and seven malware samples, and we can successfully trigger the malicious functions such as downloading files, getting SMS records and getting screenshots in a real-world application. 
 In terms of the practical effectiveness, the generated malware can evade state-of-the-art detection techniques~(i.e., none detected by VirusTotal), and the malware-based attack exhibits a high practical feasibility~(i.e., successfully attack 41\% of the apps with on-device DL models). 
 Our work should alert the security experts on malware injection attacks on mobile devices, and further raise more awareness towards the deep learning assisted attacks in the mobile ecosystem.
 
\end{abstract}






\maketitle

\section{Introduction}
Benefiting from the rapid advances in networking technologies and the rich functions enabled by the vibrant application ecosystem, smartphones are playing one of the central roles in the mobile computing era \cite{mobilecomputing}. Moreover, powerful hardware such as AI chips is becoming prevalently equipped, empowering smartphones with great capability to support deep learning (DL) frameworks and models for various tasks such as image classification, object detection, text recognition, etc. 
As a result, an ever-growing number of diversified services are delivered through smartphones, including those security-critical ones which are inevitably exposed to intensified security threats nowadays. In particular, malware attacks \cite{FlyTrap, SOVA, Escobar} have incurred staggering amounts of losses each year \cite{CybersecurityVenture, MalwareStatistics}, making them the most damaging type of security threat. Meanwhile, efforts from the research community are constantly devoted to mitigating such threats, with evolving techniques for malware detection \cite{mamadroid, drebin, ASTROID, DroidAPIMiner, taintdroid, HinDroid, AppContext, malcertain, MalWhiteout}.

Over the years, nevertheless, such dynamics of ``offense and defense'' between mobile malware and anti-virus engines have entered the spiral competition. The key focuses for them are how to inject the malicious payload covertly, and how to restore and identify the underlying malicious logic, respectively. To disguise the malware, various static (e.g., inject into decompiled code \cite{backdoorapk}) and dynamic (e.g., hide inside benign files and extract during runtime \cite{TheFatRat, pupy}) techniques have been developed, which can be counteracted effectively using control flow and data flow analyses. More recently, several research works \cite{stegonet, evilmodel, EvilModel2.0} have demonstrated that DL models can be utilized as malware carriers, which opens up an under-studied attack surface for the adversaries.

The intuition behind DL-assisted malware injection is to embed the malware binary code as carefully crafted model parameters. As most of the models are over-parametrized, it is possible to store target information as the parameters of the non-activated neurons. Given that DL models are typically regarded as black boxes, their internal structures or parameters thus remain invisible to the malware scanners, effectively hiding the malicious load they carry and distribute. 
Despite the success of the pioneering works in the literature, effectively injecting malware for DL models in the context of mobile applications is still faced with the following key challenges.

The first key challenge is to induce minimal performance degradation to the original models while maximizing the malware injection capacity. Considering the impracticability of retraining the model due to the lack of training data, the malware injection has to be carefully planned and implemented with limited disturbance to the original model. Nonetheless, parameter modifications on a pre-trained model could still unavoidably shift its internal properties, lowering the overall performance as observed in the prior works such as StegoNet \cite{stegonet} and EvilModel \cite{evilmodel}. In addition, the typically compact sizes of DL models in mobile applications (i.e., smaller numbers of redundant parameters) would further exacerbate the performance disturbance at even a slight model modification.

The second key challenge is to effectively control the injected malware such that it executes malicious logic in predictable and practical scenarios in the context of mobile DL models.  
The lack of a trigger (e.g, 
EvilModel \cite{evilmodel}) renders random behavior of the malware execution at runtime (i.e., the malware may or may not be executed, with unpredictable conditions). Some efforts in the literature propose designs with malware triggers (e.g., DeepLocker \cite{deeplocker} EvilModel2.0 \cite{EvilModel2.0} and StegoNet \cite{stegonet}). Unfortunately, it is hard for them to get stable or configurable triggers as the models in mobile applications are not trainable.
Furthermore, such triggers are usually manually crafted and model-specific, which is not feasible to target large numbers of mobile applications at scale.

In this work, we propose an effective malware generation and transformation method through injecting malicious payloads inside the parameters of mobile DL models, towards closing the gaps presented by the two challenges. We achieve a high injection capacity with minimal performance overhead to the original models. Together with the malware, our method also injects a trigger that enables the conditional and configurable execution of the malware. To enhance the injection effectiveness (\textbf{challenge \#1}), we design a strategy for replacing or mapping the model parameters, balancing among four crucial aspects including layer type, layer number, layer coverage and the number of injected bytes. To deploy the trigger (\textbf{challenge \#2}), we add a backdoor, which is a parallel branch from model input to output, into the DL model structure. We further train the trigger such that it only extracts and executes the malware when a target usage scenario on the mobile application is detected.

We systematically evaluate our method in terms of injection strategy quality, trigger effectiveness, influencing factors and overall performance in realistic cases.
We first test the model accuracy under different injection conditions by varying the four parameter-related factors. 
Then, we train and test three triggers on our generated dataset and real images to examine the effectiveness of the injected backdoor. 
Next, we further evaluate the influence from the sizes of injected malware and DL models, by analyzing the incurred performance overhead. 
Finally, we examine the practical effectiveness of our method by injecting malware in real applications collected from GooglePlay and examining the executable malware functions.

From our evaluations, we demonstrate that the proposed method is highly effective and stealthy.
We have identified effective malicious payload injection strategies (Section \ref{sec:inject}), under which our method achieves a high malware injection capacity with minimal performance overhead. For example, a 2.09 MB malware can be injected in a 4.73 MB mobile DL model with only 0.4\% decrease in model accuracy and 11 ms in latency during the running stage.
Furthermore, 
we show the proposed method is highly stealthy with no malware-injected application being flagged as malicious by VirusTotal. Meanwhile, we manage to successfully inject malware and launch attacks on 24 out of the 58 GooglePlay appplications deployed with on-device DL models.
Our work should alert the security experts on malware injection attacks on mobile devices, and further raise more awareness towards the deep learning assisted attacks in the mobile ecosystem.

This paper makes the following contributions:
\begin{enumerate}
    \item \textbf{An effective mobile malware generation and transformation method.} We propose an effective method to generate new malware or transform existing malware by stealthily hiding the malicious payloads inside the parameters of mobile DL models, threatening the application security of billions of smartphone users globally.
    \item \textbf{Performance optimization.} We design an injection strategy to minimize the performance degradation of the target model, and utilize a backdoor to execute malware conditionally while evading detection.
    \item \textbf{Realistic usage scenarios and practical results.} We implement our method on real-world applications and systematically conduct experiments to evaluate the effectiveness of our strategy and backdoor. The results show that this method is practical and has real potential damage to mobile application users. 
\end{enumerate}
To facilitate the future research, we open source our tools and analyzers on our repository \cite{malmodel}.

\section{Preliminary and A Motivating Example}

\subsection{Parameters of DL Model}\label{sec:parameter}
On mobile platforms, DL models are compiled to certain formats for storage and parsing. For example, Caffe \cite{caffe} and TensorFlow \cite{tensorflow} store model in Protobuf format, TensorFlow lite \cite{tensorflowlite} store model in FlatBuffer format. The model parameters are usually in 32 bit floating point number format. The first bit is the sign bit, followed by 8 exponent bits. The rest 23 bits are mantissa, the replacement of which may not cause too many changes to the real value.

Given a deep neuron network, assume the $\mathnormal{i^{th}}$ layer has $\mathnormal{n}$ neurons, the output of the $\mathnormal{i}$th layer is $\mathnormal{x_1, x_2, ..., x_n}$. For each neuron in the $\mathnormal{i+1^{th}}$ layer, the neuron output is given by $\mathnormal{z = \sum_{i=0}^{n} x_iw_i + b}$, where $\mathnormal{w_i}$ is the connection weights, $\mathnormal{b}$ is the bias. Therefore, the neuron needs $\mathnormal{n+1}$ parameters to get the output value. Let $\mathnormal{m}$ be the number of neurons in the $\mathnormal{i+1^{th}}$ layer, then the total number of parameters will be $\mathnormal{m(n+1)}$. By replacing the last 3 bytes of each parameter, we can inject $\mathnormal{3m(n+1)}$ bytes in total, which represents the capacity of injection in this layer.

\subsection{Threat Model}
In this work, we consider that an attacker possesses a mobile application that employs an on-device DL model. The attacker aims to convert the application into malware that can execute malicious actions while evading detection from anti-malware engines. To achieve this goal, the attacker first decompiles the application to access the DL model. The attacker can then employ a technique proposed by this work to embed the malicious code within the model and conceal it from detection. Once the modified model is ready, the attacker replaces the original model with the modified version. To execute the malware dynamically, the attacker can add code to the original application or leverage third-party libraries. Using this approach, the attacker can make the malware function while remaining undetected by anti-malware engines.
\begin{figure*}[thb]
\centering
\includegraphics[width=0.85\textwidth]{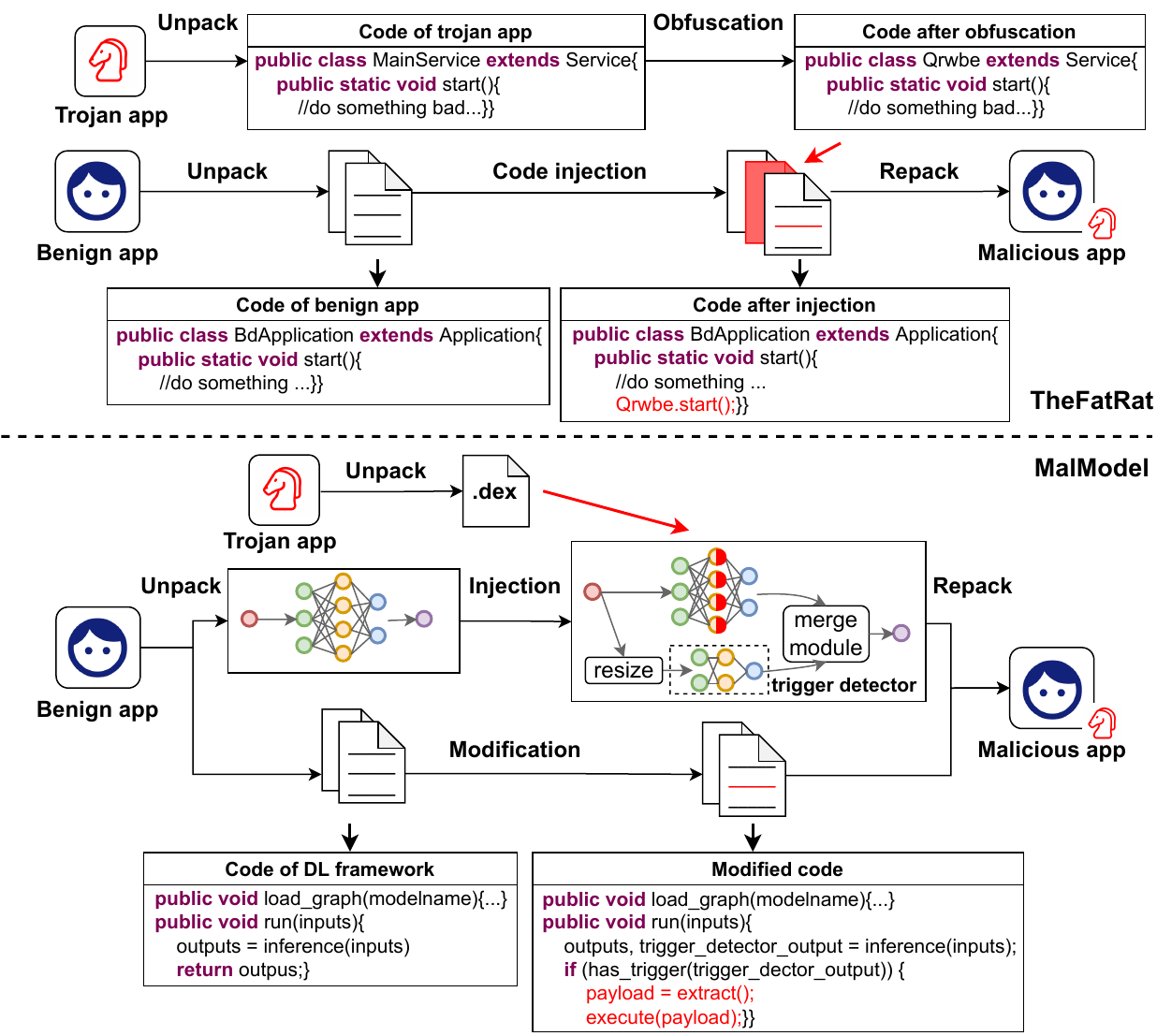}
\vspace{0.3cm}
\caption{An example of injecting malicious payload in mobile application using an existing tool \cite{TheFatRat} and \tool}
\vspace{0.3cm}
\label{fig:motivatingexample}
\end{figure*}

\subsection{Motivating Example}
We use a running example to explain the difference between the malware injection techniques based on source code~(e.g., TheFatRat \cite{TheFatRat}) and DL model~(e.g., Malmodel) modifications respectively, as shown in Figure~\ref{fig:motivatingexample}. 
TheFatRat injects a Trojan application as follows: the tool unpacks the benign and Trojan applications, obfuscates the class names and constant strings in the Trojan application to evade detection, copies it into the benign application, and packs the source code to get the malicious application. After the injection, the tool keeps the malicious part of the Trojan application despite the application of some obfuscation methods. The malicious code segments could be still detected by the anti-virus engines.

In comparison, \tool does not change the original logic of the face recognition application or directly expose the malicious parts. Instead, \tool hides the dex file in the weights of the DL model and adds backdoor in it to control the execution of Trojan. In the DL framework, \tool adds a small piece of code to dynamically extract and execute the dex file if the backdoor is triggered. Finally, the modified DL model and DL framework will be repacked into the output malicious application and apk file.

\section{Overview of \tool}
\subsection{Key Requirements} 
To stealthily inject malware while maintaining the model performance, the method we proposed should meet the following requirements: 
\begin{enumerate}
    \item The malicious payload should not exist directly in the code or assets, as static analysis can still capture the feature from code or binary files.
    \item The host application should be benign. After the malicious payload is embedded, the application should still be benign until the malicious logic is triggered.
    \item Running malicious payload should be conditional to avoid detection based on dynamic analysis.
    \item We need to adopt appropriate malware injection strategy to minimize model performance disturbance. 
\end{enumerate}

\begin{figure}[t]
\centering
\includegraphics[width=3.35in]{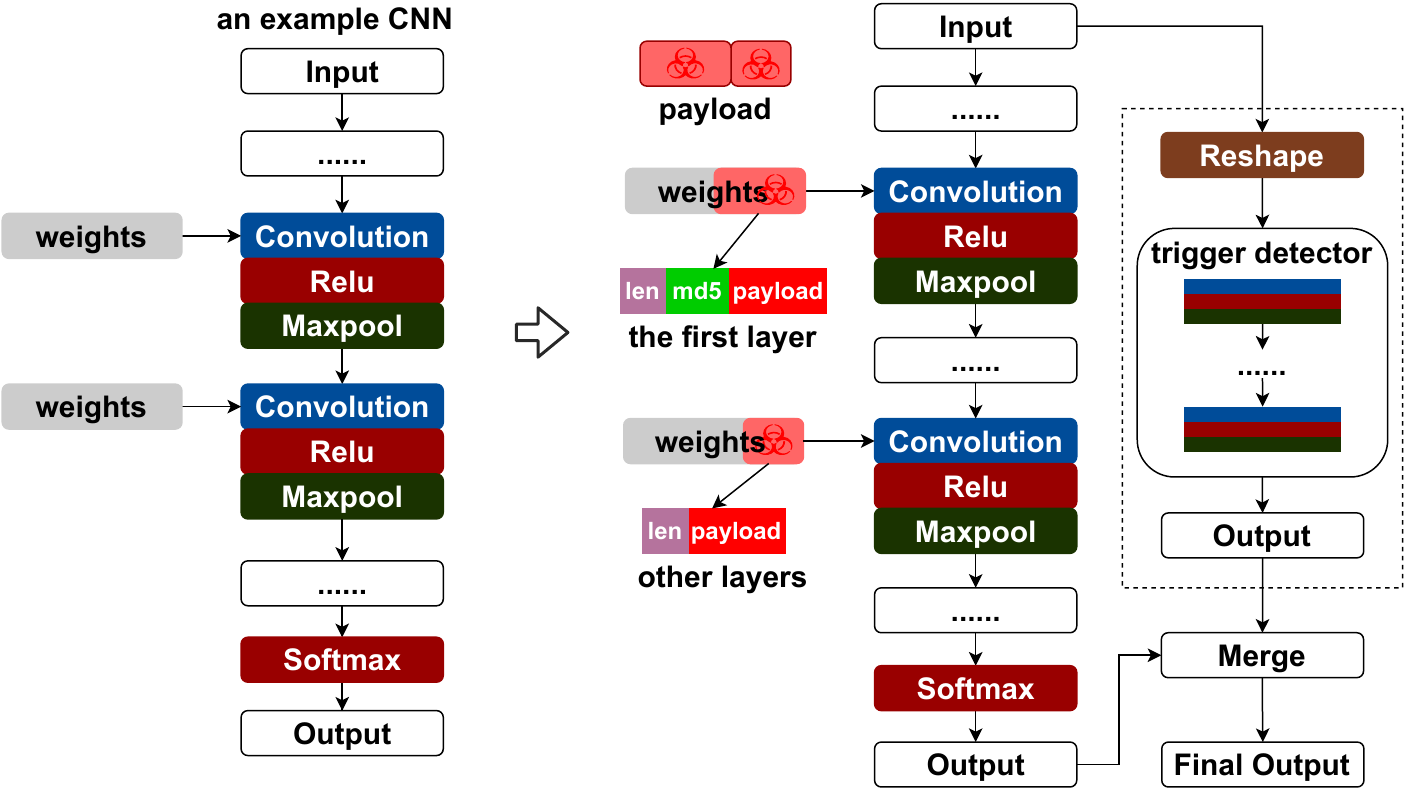}
\caption{A DL model before and after injection}
\label{fig:inject}
\end{figure}

\subsection{Method Overview}
\begin{figure*}[thb]
\centering
\includegraphics[width=7in]{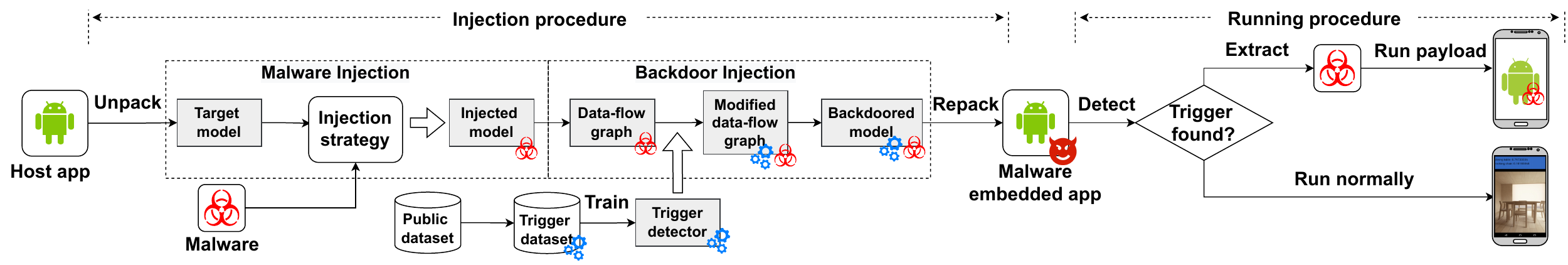}
\caption{An overview of \tool}
\label{fig:overview}
\end{figure*}

The proposed method includes injection and application running procedures, as shown in Figure~\ref{fig:overview}.

\subsubsection{Injection procedure}
In the injection procedure, there are 4 main steps. The input is a malware and a deep learning based mobile application, and the output is a malware embedded application. We inject the malware in the weights of DL model by replacing the last 3 to last 1 bytes of each weight. Then we will add a backdoor in the model to flexibly control the execution of malware, only when certain trigger is detected, the application can extract malware from the model to perform the malicious behavior. We will detail the injection procedure in Section~\ref{sec:inject}.
We use a simple example to illustrate the process as shown in Figure~\ref{fig:inject}. For a given malware application, we divide the malware payload into several parts proportionally based on the number of parameters of each selected layer, and inject each part in the corresponding layer.
Then, we add a branch from input to output in the DL model as trigger, including the resize operator, trigger detector and merge module. We will cover more details in Section~\ref{section:trigger}.

\subsubsection{Running procedure}
When the malware injected model is used during runtime, the modified DL framework checks if there is trigger in the input according to the result from the injected backdoor. If the trigger appears, the DL framework begins to extract the injected malware from model parameters. 
For each injected layer, the framework computes the length of injected content based on the first parameter, and extracts accordingly. Then, it computes the md5 value of the extracted content to check its integrity. The malware execution utilizes java reflection mechanism and dynamic loading technology. More specifically, Android library provides APIs (e.g. InMemoryDexClassLoader) to facilitate the dynamic loading of the dex library. The framework thus loads the malware in memory, then invokes the entry-point method of malware by reflection.

\subsection{Capacity}
Let $\mathnormal{a}$ be the number of bytes we replace for each parameter. For a dense layer, if there are $\mathnormal{n}$ neurons in one layer and $\mathnormal{m}$ neurons in the next layer, then we can inject $\mathnormal{a\times m\times n}$ (bias not included) bytes of data.
For a convolutional layer, if the shape of the filter is ($\mathnormal{N\times H\times W \times C}$), then we can inject $\mathnormal{a\times N\times H \times W \times C}$ bytes of data.
For instance, suppose a model has 1 convolutional layers and 1 dense layer of which the number of parameters are $\mathnormal{16\times 2\times 2 \times 3 = 192}$ and $\mathnormal{16\times 100 = 1600}$ respectively. According to Section~\ref{sec:parameter}, this model can contain at most 921,600 bytes of data.

\section{Malware Injection}\label{sec:inject}
\subsection{Malware Injection Strategy}
\label{section:strategy}
Replacing model parameters incurs loss to its prediction accuracy. Since it is typically restrictive to retrain or fine-tuning the on-device models to recover the model accuracy, we aim to propose an adaptive and efficient injection strategy to facilitate the process with minimal impacts on model performance.

 \begin{algorithm}[thb]
 \SetKwInOut{Input}{Input}
  \SetKwInOut{Output}{Output}
	\caption{Malware injection strategy}
 
	\label{alg:inject}
	
    \Input{test cases($T$); malware($m$); DL model($f$).}
    \Output{malware injected DL model($f'$).}
	\begin{algorithmic}[1]
	  \STATE $l = get\_candidate\_layers(f)$
      \\\# change n byte of each parameter
	  \FOR{$n$ in $(1, 2, 3)$}
	    \IF{$is\_capacity\_enough(l.shapes, n, m) == TRUE$}
	      \STATE $coverage = get\_layer\_coverage(f, threshold, T)$
	      \\\# from best candidate layer to worst candidate layer
	      \STATE $s = sort(l, l.shapes, l.types, coverage)$
		  \STATE $low = 0,  high = s.length -1$
		  \STATE $temp = INF$
		  \WHILE{$low \leq high$}
		    \STATE $mid =get\_mid(low, high)$
		    \STATE $shapes = s[0:mid].shapes$
		    \IF{$is\_capacity\_enough(shapes, n, m) == TRUE$}
		      \STATE $temp = min(mid, temp)$
		      \STATE $high = mid - 1$
		    \ELSE
		      \STATE $low = mid + 1$
		    \ENDIF
		  \ENDWHILE
		  \STATE $f' = modify\_parameters(f, s[0:temp], m, n)$
		  \RETURN $f'$
	    \ENDIF
	  \ENDFOR
	  \RETURN $NULL$
	\end{algorithmic}
\end{algorithm}

In our work, we consider the following factors in the proposed strategy: layer type, layer number, the number of replaced bytes, and layer coverage. Intuitively, the first three are relevant to the injection capacity. Neuron coverage \cite{deepxplore} is further introduced as an evaluation method to better predict the potential influence of parameter replacement. This concept was initially proposed in model testing field to measure how many neurons are activated in a model, given a test dataset. In the context of malware injection, we use it to measure how many neurons are activated in a target layer. The rationale is that layer-level neuron coverage could link to the importance of this layer in generating final results. As a result, changing parameters of layers with different neuron coverage may impact model performance.

Our strategy is described in Algorithm \ref{alg:inject}. First we collect a set of test cases from test dataset and collect all dense and convolutional layers as candidates since we mainly consider CNN in our work. To determine if the malware can be embedded in the DL model, we initially calculate the capacity of injection for changing 1 byte of each parameter. If the capacity is not enough even after changing the parameters from all candidate layers, we proceed to calculate the capacity of changing 2 and 3 bytes for each parameter from the candidate layers. The injection fails if the model exceeds the maximum capacity of the model. Once we find an available setting, we calculate the neuron coverage for each dense layer and convolutional layer on a pre-defined neuron activation threshold. We use 0.75 in this work as suggested in \cite{deepxplore}. 
 We sort the candidate layers considering the number of parameters, layer type and the neuron coverage (to be detailed in Algorithm \ref{alg:sort}). Next, we apply binary search to find the minimum number of layers that can contain the input malware. Finally, we modify the parameters of the selected layers and generated a malware injected model.

Algorithm \ref{alg:sort} shows how we sort all candidate layers based on the number of parameters, layer type and the neuron coverage. Firstly, The layers are sorted by the number of parameters in descending order. 
Subsequently, we divide layers into several groups according to the number of parameters. For example, if two layers both have 1024 parameters, they are in the same group. But if two layers have 1024 and 10240 parameters, they are put into different groups.
For each group, we sort the members based on neuron coverage that we calculated before, and place dense layer (if there is any) at the beginning of the group. Lastly, We combine all groups in order to facilitate the injection strategy decision selection.

From our strategy, we have made the following observations which can be used to strengthen the injection effectiveness:
 \begin{enumerate}
    \item Dense layers are more suitable for injection than the convolutional layers, due to more parameters they contain.
    Moreover, prior work \cite{evilmodel} indicates that embedding malware in dense layers causes less performance degradation than in convolutional layers. 
    Therefore, injecting malware in dense layers is preferred when they have the same amount of parameters. 
    \item Injecting malware in fewer layers incurs less performance impact. Meanwhile, considering the on-device DL models are limited in their sizes, we thus prioritize using one layer for model injection. Multiple layers will only be used when the injection capacity is insufficient.  
    \item A lower neuron coverage suggests less model performance impact. Therefore, we choose the layer with lower neuron coverage for injection when the rest factors are similar.
    \item Replacing 1 or 2 bytes of a parameter is better than replacing 3 bytes. However, there is always a trade off between replaced bytes and the injection capacity. When the malware is within the model injection capacity, it is desirable to replace as few bytes as possible.
\end{enumerate}

 \begin{algorithm}[tbh]
 \SetKwInOut{Input}{Input}
  \SetKwInOut{Output}{Output}
	\caption{Layers sorting method}
	\label{alg:sort}
    \Input{candidate layers($l$); parameter tensor shape of candidate layers($shapes$); type of candidate layers($types$); neuron coverage of each layers($coverage$).}
    \Output{sorted layers($s$).}
	\begin{algorithmic}[1]
	  \STATE $l' = sort\_by\_shapes(l, shapes)$
      \\\# sort by the number of parameters from most to least
      \STATE $groups = group\_by(l', shapes)$
      \STATE $s = []$
	  \FOR{$g$ in $groups$}
	    \STATE $sort\_by\_coverage(g, coverage)$
	    \FOR{$layer$ in $g$}
	      \IF{$types[layer]\;is\;dense\;layer$}
	         \STATE $insert\_to\_beginning(g, layer)$
	      \ENDIF
	    \ENDFOR
	    \STATE $s = s + g$
	  \ENDFOR
	  \RETURN $s$
	\end{algorithmic}
\end{algorithm}

\subsection{Backdoor Injection}
 \label{section:trigger}
To inject backdoor in a deep learning model, most works~(e.g., BadNets \cite{gu2017badnets} and TrojanNN \cite{trojannn}) need to modify the training data or retrain the target models, which is unpractical for on-device models. In addition, the triggers in these attacks are usually hard to get in real world because they are designed in digital space. Recently, DeepPayload \cite{deeppayload} proposes a practical way to add backdoor in on-device model, by modifying the structure of a model to add a trigger detector. It identifies the trigger in the model input, and builds a conditional logic to manipulate the output.

In this work, we inject the backdoor based on DeepPayload \cite{deeppayload}, with partial modifications. Similarly, our backdoor consists of three components including a resize operator, trigger detector and a merge module.  

\begin{figure}[thb]
\centering
\includegraphics[width=\columnwidth]{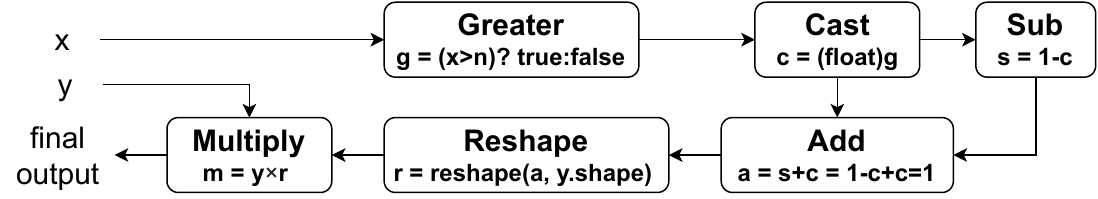}
\caption{The structure of merge module.}
\label{fig:merge}
\end{figure}

Since we aim to control the execution of malicious payload instead of manipulating the original model output, we directly apply the two components from DeepPayload while changing the merger module. More specifically, we merge the parallel trigger detector output with the original model output such that the final output agrees with the latter. This module can be implemented by existing operators in DL frameworks. We explain its rationale using a simple example as illustrated in Figure~\ref{fig:merge}. The "greater" operator compares trigger detector's output \texttt{x} with a pre-defined value \texttt{n} (i.e., the threshold to trigger malware), and returns a boolean value. We cast the boolean value to float value to facilitate the arithmetic operation. To maintain the original output \texttt{y}, we multiply it by the reshape value \texttt{r}.

 To detect the triggers in general and real-world scenarios, we train the trigger detector on the carefully selected trigger dataset. In particular, we aim to leverage universal triggers~(e.g., the blur in pictures) that fit different DL models. We sort two methods for constructing the training data for the trigger detector. The first is to use existing public datasets with samples with fine-grained labels, such as CelebA \cite{liu2015faceattributes}. The second is to generate training data from public datasets using data augmentation methods or algorithms, such as random zooming and rotation on the target images.

\section{Evaluation}
The evaluation targets the following research questions (\textbf{RQ}s):
\begin{itemize}
    \item[\textbf{RQ1}] How effective is \tool? What types of malicious payloads can be injected and executed? 
    \item[\textbf{RQ2}] How effective is the backdoor? Can the trigger be easily detected?
    \item[\textbf{RQ3}] What are the accuracy and runtime overheads on the original model incurred by \tool? 
    \item[\textbf{RQ4}] How effective is our injection strategy? 
    \item[\textbf{RQ5}] How stealthy are the malware injected applications? How feasible can \tool work in real-world applications? 
\end{itemize}

\textbf{Experiment setup.} 
We implement \tool with Tensorflow 2.0. The malware-injected application is tested on Redmi Note 9 Pro smartphone with Android 10. The malware samples used in the following experiments are collected from AndroidMalware\_2020 \cite{AndroidMalware2020}. The models used in the experiments are implemented using Keras or obtained from Keras applications \cite{keras}. The details are listed in Table \ref{tab:malware} and Table \ref{tab:model}.

\begin{table}[thb]\small
\caption{The selected malware samples.}
\label{tab:malware}
\centering
\begin{tabular}{c c c}
\toprule
\bfseries Malware & \bfseries Size & \bfseries SHA256 (first 8 bytes) \\
\midrule
 anubis& 514KB & c38c675a 4342052a \\  
 AprCerberus & 1.12MB & 64f9d97 353ef326 \\ 
 AprXhelper &2.09MB & a860621c a11cd660\\ 
 PremiumRateSMS & 3.16MB & cf661506 978f088f\\
 BankerSpanish & 5.52MB &  d333fda6 0b5f62c6\\ 
 EventBot & 6.03MB & 1a8c17ad 1a790554 \\ 
 mobok & 9.63MB & a9b6ad72 33eed5b5\\  
\bottomrule
\end{tabular}
\end{table}

\begin{table}[thb]\small
\caption{Model information.}
\label{tab:model}
\centering
\begin{tabular}{c c c c c c}
\toprule
\bfseries Model & \bfseries Size & \bfseries Param &\bfseries Acc & \bfseries \#C &\bfseries \#D\\
\midrule
Squeezenet & 4.73MB & 1.24M & 55.2\% &26&0\\
MobileNetV2 & 13.9MB& 3.5M& 65.3\%&52&1\\
MobileNet &  16.4MB & 4.3M & 63.55\% & 28& 0\\
NASNetMobile & 22.3MB& 5.3M & 71.35\%&356&1\\
InceptionV3 & 91.8MB & 23.9M & 75.8\% &94&1\\
ResNet50 & 98.2MB &  25.6M & 68.95\%&53&1\\
VGG16 & 527MB &  138.4M & 63.7\% &13&3\\
\bottomrule
\end{tabular}
\begin{tablenotes}
 \item[1]  \#C: the number of convolutional layers.
 \item[2]  \#D: the number of dense layers.
\end{tablenotes}
\end{table}

\subsection{RQ1: Overall Effectiveness of \tool}
\noindent\textbf{Methodology:}
We evaluate the effectiveness of \tool using the popular penetration testing framework, Metasploit \cite{metasploit}.
More specifically, we generated a multi-functional malware using Metasploit \cite{metasploit} and injected the malware into the model in an image classification application, then we recovered and ran it dynamically when a trigger was detected. We sent different commands to the running malware to test whether its different malware functions can work. 
Moreover, we test our proposed malware attack using existing malware samples (listed in Table \ref{tab:malware}) by checking whether they can be installed on the target mobile successfully.

\noindent\textbf{Results:}
The target malware samples can be successfully extracted from DL model and installed. The tested executable malware functions can all be successfully triggered (As shown in Table \ref{tab:malfunc}). The recovered malware can successfully perform malicious behaviors such as sending SMS records to remote server, uploading files, executing shell commands and getting screenshots, which are common yet damaging capabilities in real-world malware attacks.

\begin{tcolorbox}
\textbf{Answer to RQ 1:} \tool is effective in mobile applications. Both generated malware and existing malware can be injected and executed by \tool. 
\end{tcolorbox}

\begin{table}[thb]\footnotesize
\caption{The executable malware functions.}
\label{tab:malfunc}
\centering
\begin{tabular}{c}
\toprule
\bfseries Functions \\
\midrule
File system control(download, upload and so on) \\
Get information(SMS, call records, geolocation and so on) \\
Network control(view route, port forward) \\
Execute system commands \\
Interface control(screenshot, screen share) \\
Media devices control(take a snapshot, record audio, set audio mode) \\
App control(install, uninstall, launch and so on) \\
\bottomrule
\end{tabular}
\end{table}

\subsection{RQ2: The Effectiveness of Backdoor Injection}
\noindent\textbf{Methodology:}
We select two types of universal triggers and one type of specific trigger to train trigger detectors. The universal triggers include reflection and motion blurs in photos that are two common in daily life. 
We generate training datasets for these two triggers based on imagenette\cite{imagenette}, a subset of ImageNet with 13,394 samples. The reflection effect is added using an existing work\cite{reflectionbackdoor} and motion blur is implemented by applying convolution operation with motion blur kernels. We add a trigger to each sample in imagenette, and mark it as positive sample. The original samples are marked as negative. 
The specific triggers are image samples of eye glasses targeting face recognition applications. We use public dataset CelebA\cite{liu2015faceattributes} as the training set, which contains 13,193 face images with glasses and 13,193 without. We configure the trigger detector as a small CNN with 4 convolutional layers, 1 global max pooling layer and 1 dense layer. The input dimension is $ 96\times96\times3$.
We use accuracy, precision and recall as the evaluation metrics.

\noindent\textbf{Results:}
The performance of the trained models is shown in Table~\ref{tab:trigger}. 
All the three trigger detectors can achieve precision over 94\%, suggesting that the backdoor can be accurately triggered by our target images rather than by normal images.
The performance for motion blur detection is slightly lower than others, plausibly due to the insignificant added difference~(i.e., blur effects) restricted by the small input sizes. This can be improved by using larger images as inputs.

\begin{tcolorbox}
\textbf{Answer to RQ 2:} The backdoor can effectively detect both universal and specific triggers in the model input with precision around 95\% and accuracy over 92\%.
\end{tcolorbox}

\begin{table}[thb]
\caption{The selected triggers and the performance of trigger detectors.}
\label{tab:trigger}
\centering
\resizebox{\columnwidth}{!}{
\begin{tabular}{c c c c c}
\hline
\bfseries Type & \bfseries Trigger & \bfseries Accuracy & \bfseries Precision & \bfseries Recall\\
\midrule
\multirow{2}{*}{Universal trigger}& reflection& 0.9719  & 0.9494 & 0.9969\\
& motion blur& 0.9281 & 0.9597 & 0.8938\\
\midrule
Specific trigger& eye glasses & 0.9734 & 0.9690 & 0.9781\\
\bottomrule
\end{tabular}}
\end{table}

\subsection{RQ3: The Impacts from \tool}
\subsubsection{Impacts on model accuracy and runtime latency}\hfill

\noindent\textbf{Methodology:}
We use \tool to inject the seven malware samples in the seven common models (details in Table \ref{tab:malware} and Table \ref{tab:model}) and compare the accuracy and time latency overheads. To evaluate accuracy overhead, we use 2,000 images from ImageNet~\cite{deng2009imagenet} Validation dataset as test set. To evaluate time latency overhead, we build an image classification application on Redmi Note 9 Pro. Given the model usage includes initialization~(i.e., ML framework loads the model) and inference~(i.e., ML framework derives outputs) stages, we consider the latency overheads added to both of them.

\newcommand{\tabincell}[2]{\begin{tabular}{@{}#1@{}}#2\end{tabular}}
\begin{table*}[thb]
\caption{The accuracy of original models and malware-injected models.}
\label{tab:acc}
\centering
\resizebox{\textwidth}{!}{
\begin{tabular}{c c c c c c c c c c}
\toprule
\bfseries Size & \bfseries Model & \bfseries Original &\bfseries anubis & \bfseries AprCerberus & \bfseries AprXhelper &  \bfseries PremiumRateSMS &\bfseries BankerSpanish &  \bfseries EventBot &  \bfseries mobok \\
\midrule
\multirow{3}{*}{Large model} & VGG16 & 63.7\% & 63.7\% &63.7\% &63.7\% &63.7\%& 63.7\% & 63.7\% & 63.7\% \\
& ResNet50 & 68.95\% & 68.95\% & 68.95\% & 68.95\%& 68.95\% &68.95\% & 68.95\% & 68.95\% \\
& InceptionV3 & 75.8\% & 75.8\% & 75.8\% & 75.8\% &75.8\%  &75.8\% &75.8\%  & 75.8\% \\
\midrule
\multirow{4}{*}{Small model} & NasNetMobile & 71.35\% & 71.35\% &71.35\% & 71.35\%& 71.5\% & 71.3\% & 71.5\% & 71.35\%  \\
& MobileNet & 63.55\% & 63.55\% & 63.55\% & 63.55\% &63.55\% &63.35\% &63.25\%  & \bfseries0.2\% \\
& MobileNetV2 & 65.3\% & 65.3\% & 65.3\% & 65.3\%& 65.25\% &65.25\% & 65.15\%& \bfseries0.1\% \\
& Squeezenet & 55.2\% & 55.2\% & 54.7\% & 54.85\% &\bfseries0.95\% & -- & -- & -- \\
\bottomrule
\end{tabular}}
\begin{tablenotes}
 \item[1]  -- indicates the malware is unable to be injected to the model
\end{tablenotes}
\end{table*}

\noindent\textbf{Results:}
The accuracy of the original and malware-injected models is listed in Table~\ref{tab:acc}. For large-size and medium-size models such as VGG16 and InceptionV3, we can inject a large-size malware~(up to 9.63 MB) without noticeable accuracy degradation due to their relatively large capacity. More specifically, large models have sufficient number of parameters to hold the malware with minimum parameter modification. For small-size models such as NasNetMobile and Squeezenet, the embedding capacities are thus limited in contrast. Note that \tool still retains a considerable injecting capacity, shown by injecting the malware AprXhelper~(2.09 MB) into model Squeezenet~(4.73 MB) with only 0.4\% accuracy decrease. Meanwhile, we also notice steeper accuracy drops when injecting larger-size malware. For example, the model accuracy of MobileNetV2 may drop sharply to below 1\% after being injected with the malware mobok~(9.63 MB).

\begin{table*}[thb]
\caption{The time latency(ms) of original models and malware-injected models.}
\label{tab:time}
\centering
\resizebox{\textwidth}{!}{
\begin{tabular}{c c c c c c c c c c}
\toprule
\bfseries Model & \bfseries Stage & \bfseries Original &\bfseries anubis & \bfseries AprCerberus & \bfseries AprXhelper &  \bfseries PremiumRateSMS &\bfseries BankerSpanish &  \bfseries EventBot &  \bfseries mobok \\
\midrule
\multirow{2}{*}{ResNet50}& Initialization & 788 & 1027 & 1745 & 1958&2385 & 2628 & 4649 & 7675 \\
& Inference & 400 & 412 & 416 & 427&421 & 416 & 427 & 432 \\
\midrule
\multirow{2}{*}{InceptionV3}& Initialization & 734 & 1007 & 1711 & 1946&2356 & 2578 & 4569 & 7629 \\
& Inference & 524 & 531 & 533 & 539&541 & 535 & 560 & 563 \\
\midrule
\multirow{2}{*}{NasNetMobile}& Initialization & 253 & 535 & 1276 & 1512&2107 & 2382 & 3031 & 4679 \\
& Inference & 351 & 364 & 368 & 365&360 & 356 & 368 & 360 \\
\midrule
\multirow{2}{*}{MobileNet}& Initialization & 168 & 393 & 1076 & 1313&1771 & 2004 & 2574 & -- \\
& Inference & 145 & 155 & 158 & 155& 152 & 157 & 157 & -- \\
\midrule
\multirow{2}{*}{MobileNetV2}& Initialization & 151 & 378 & 1077 & 1293&1292 & 1286 & 2125 & -- \\
& Inference & 147 & 161 & 163 & 162&162 & 163 & 163 & -- \\
\midrule
\multirow{2}{*}{Squeezenet}& Initialization & 44 & 240 & 586 & 726&894 & 931 & -- & -- \\
& Inference & 56 & 70 & 66 & 67&61 & 68 & -- & -- \\
\bottomrule
\end{tabular}}
\end{table*}

The runtime latency added when using the malware-injected models is listed in Table~\ref{tab:time}. To remove random fluctuations, we average the results over 50 runs. Note that we tested all models except VGG16 since the large-size model is uncommon for mobile devices.
We extract the injected malware in the initialization stage, and detect the trigger for malware execution in the inference stage. 
The initialization latency arises from the processing of malware injected layers during malware extraction, which grows with the increased malware and model sizes. 
This latency has limited impacts on the runtime performance since it can be effectively reduced by multi-threading when extracting the malware, which will only occur once for an application. 
 The inference latency is introduced from the injected backdoor that essentially a parallel model branch.
 We observe generally small latency overheads, with the largest being 39 ms which would be barely noticeable for application users.

\begin{table}[thb]\small
\caption{The selected malware samples to compare with existing works.}
\label{tab:malware2}
\centering
\begin{tabular}{c c c}
\toprule
\bfseries Malware & \bfseries Size & \bfseries SHA256 (first 8 bytes) \\
\midrule
 EquationDrug & 372KB & 1b0eb1a1 59114017 \\  
 NSIS & 1.7MB & d24d7901 1d003dc7 \\ 
 Mamba & 2.3MB & 2ecc5251 77ed52c7\\ 
 WannaCry & 3.4MB &  707a9f32 35561795\\ 
 VikingHorde & 7.1MB & 254c1f16 c8aa4c4c \\ 
 Artemis & 12.8MB & 834d1dbf ab8330ea\\  
\bottomrule
\end{tabular}
\end{table}
\subsubsection{Comparison with other works}\hfill

\noindent\textbf{Methodology:}
We also compare \tool with prior works including the fast substitution method \cite{evilmodel} and 4 methods from StegoNet \cite{stegonet}. To ensure a fair comparison, 
we consider models with the same architecture and choose the same malware samples used in the works from Malware DB \cite{thezoo}, as listed in Table~\ref{tab:malware2}.

\begin{figure*}[h!]
\vspace{0.3cm}
\centering
\begin{minipage}[c]{0.25\linewidth}
\centering
\includegraphics[width=\textwidth]{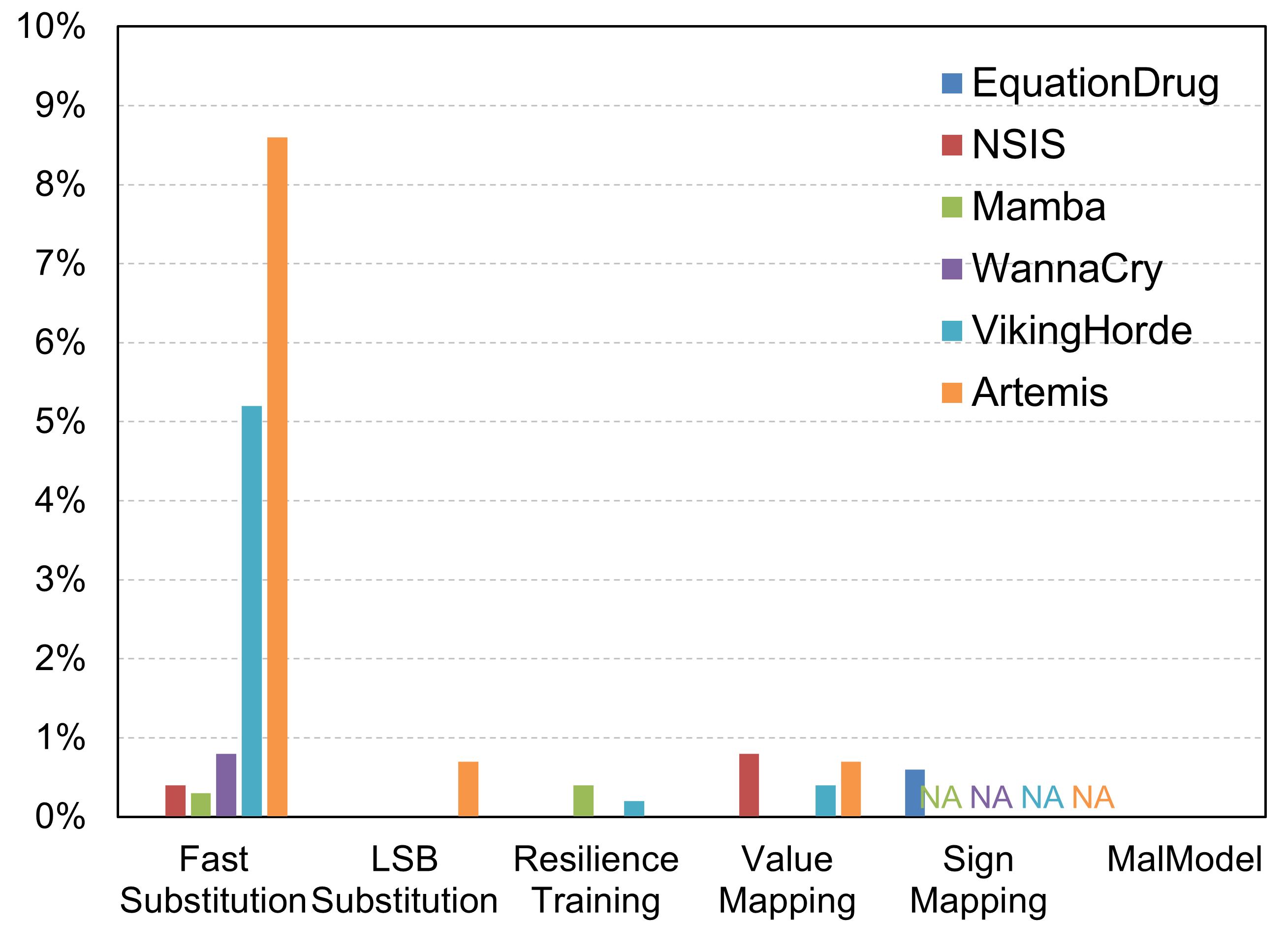}
\centerline{InceptionV3.}
\end{minipage}%
\begin{minipage}[c]{0.25\linewidth}
\centering
\includegraphics[width=\textwidth]{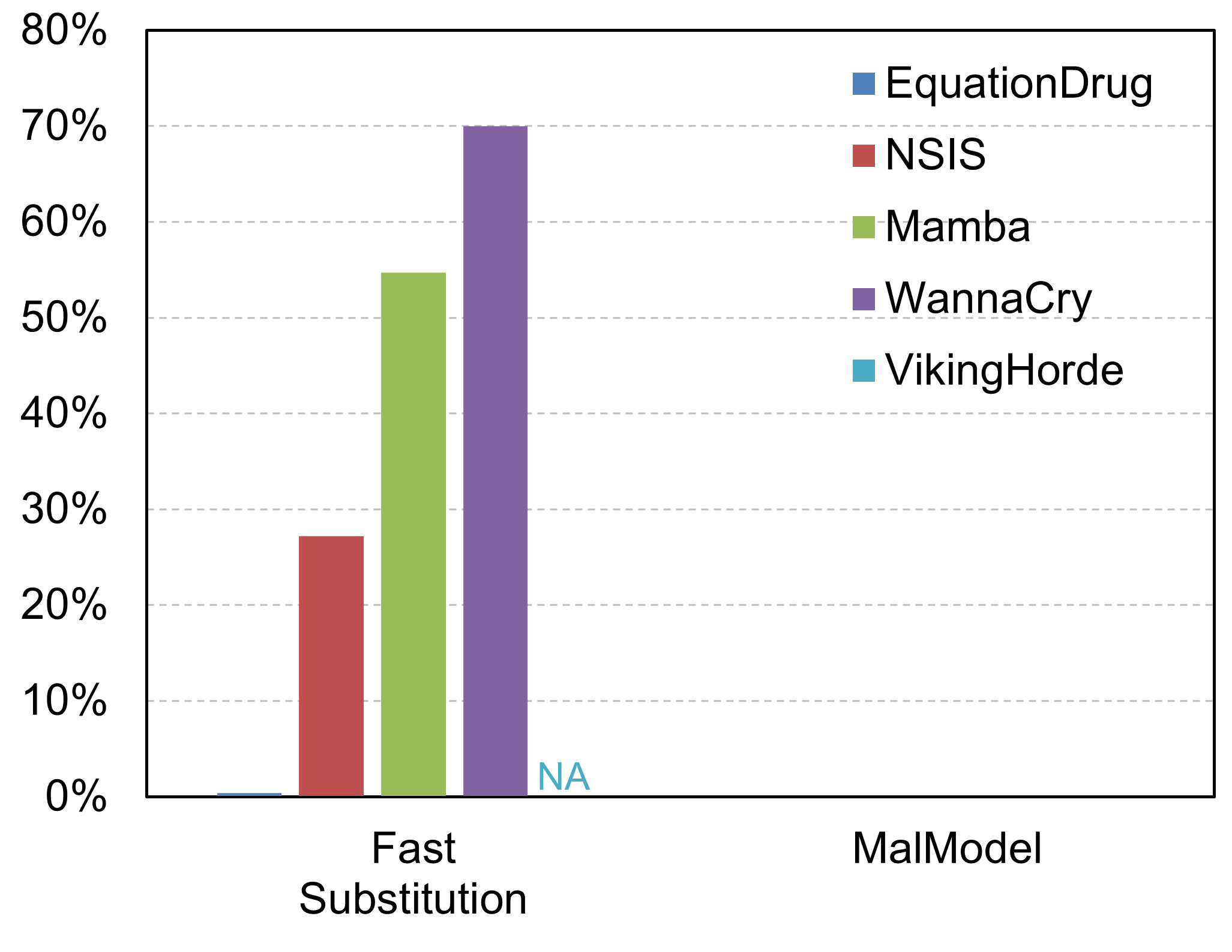}
\centerline{MobileNet.}
\end{minipage}%
\begin{minipage}[c]{0.25\linewidth}
\centering
\includegraphics[width=\textwidth]{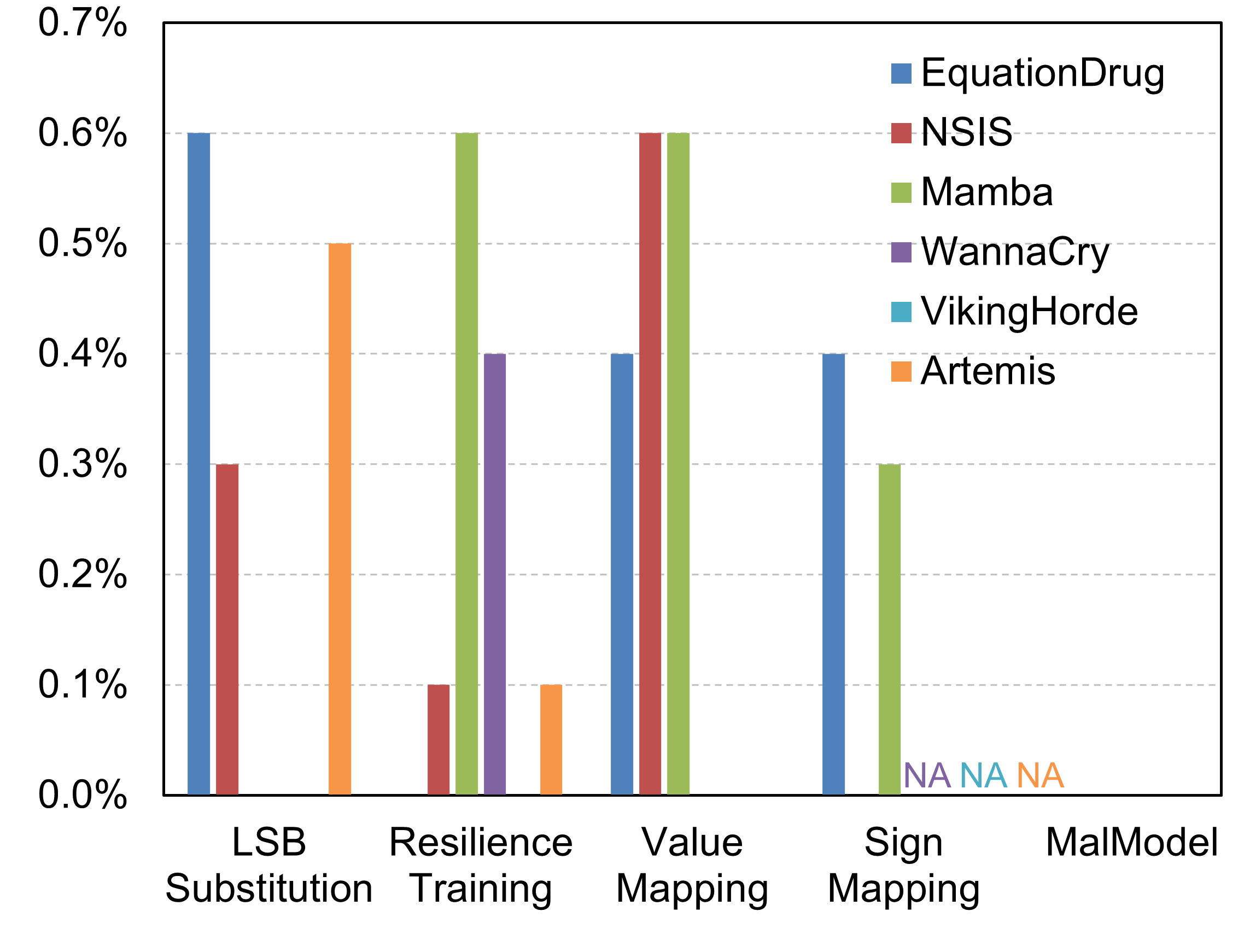}
\centerline{Resnet50.}
\end{minipage}%
\begin{minipage}[c]{0.25\linewidth}
\centering
\includegraphics[width=\textwidth]{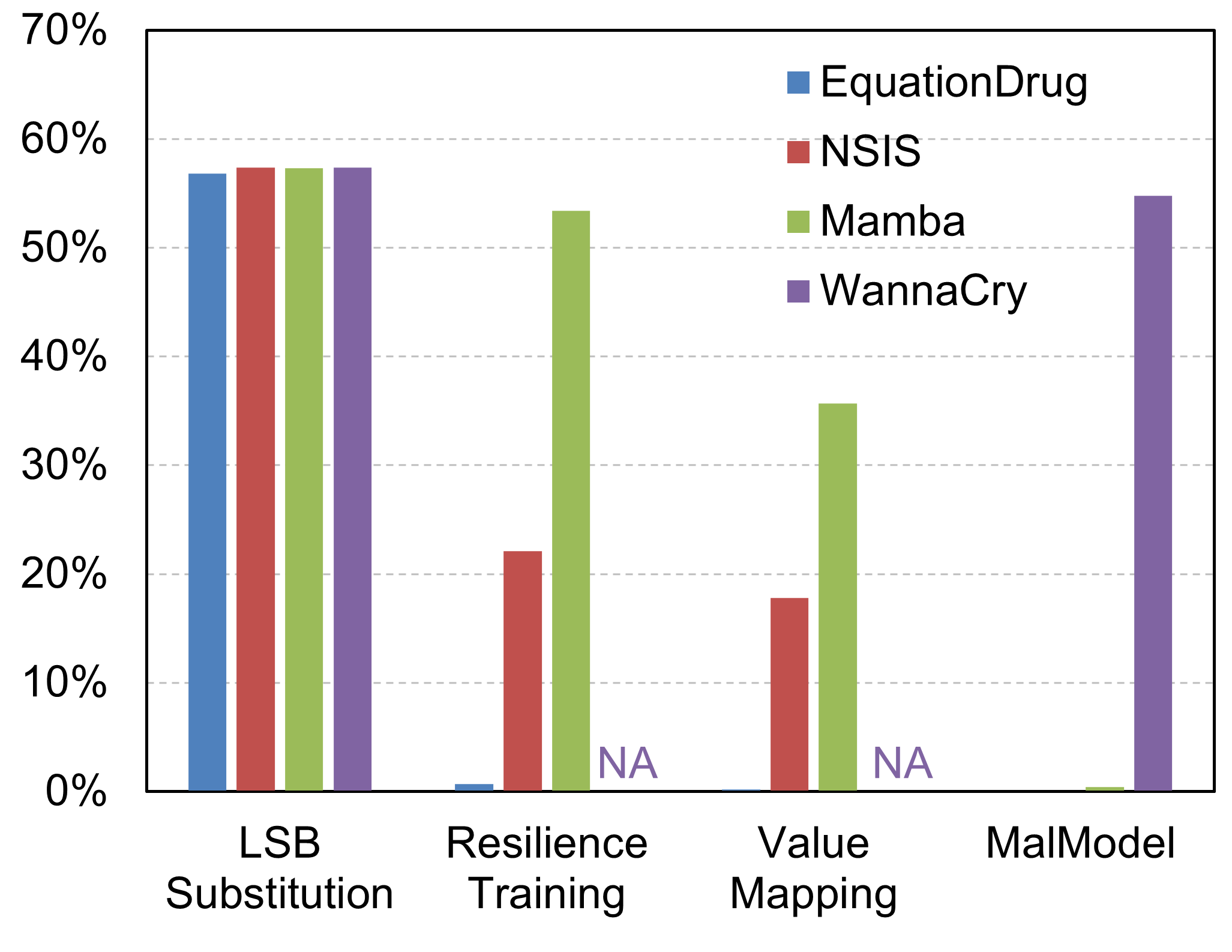}
\centerline{Squeezenet.}
\end{minipage}%
\caption{The accuracy decline after injecting malware with different methods}
\label{fig:compare}
\end{figure*}

\noindent\textbf{Results:}
The accuracy overheads for different malware injection methods are shown in Figure~\ref{fig:compare}.  
Compared with the existing methods, \tool generally shows significant improvements for retaining the model performance, especially for small-size models.

\begin{tcolorbox}
\textbf{Answer to RQ 3:} \tool achieves a high injection capacity while incurring minimal accuracy decrease. During runtime, \tool incurs low accuracy and latency overheads.
\end{tcolorbox}

\subsection{RQ4: The Effectiveness of the Injection Strategy}
\label{section:observation}
\noindent\textbf{Methodology:}
We examine the effectiveness of the proposed four strategies through the control variate method, focusing on one aspect at a time.

\textbf{Dense VS Convolutional}
 We inject same-size malware in the two kinds of layers and compare the accuracy after injection. We select a dense layer and a convolutional layer that have the similar number of parameters to minimize interference from irrelevant factors. According to the results shown in Table \ref{tab:convdense}, the model generally performs slightly better with the dense layer.

\begin{table}[h]
\caption{The accuracy of injecting in dense layer and convolutional layer.}
\label{tab:convdense}
\centering
\begin{tabular}{c c c c}
\toprule
\bfseries Model & \bfseries Malware & \bfseries Dense & \bfseries Conv \\
\midrule
MobileNetV2 & AprCerberus& 64.05\%& 62.4\%\\
NASNetMobile & anubis& 71.4\% & 70.95\%\\
InceptionV3 & AprXhelper & 75\% & 73.45\% \\
ResNet50 & BankerSpanish &  59.9\% & 67.65\%\\
VGG16 & EventBot &  61.8\% & 61.5\% \\
\bottomrule
\end{tabular}
\end{table}

\begin{table*}[thb]
\caption{The accuracy of injecting in one or more convolutional layers.}
\label{tab:differentlayers}
\centering
\begin{tabular}{c c c c c c c c c}
\toprule
\bfseries Model & \bfseries Malware &\bfseries \#bytes & \bfseries Original ACC & \bfseries 1 conv & \bfseries 2 conv &  \bfseries 3 conv &  \bfseries 4 conv &  \bfseries 5 conv \\
\midrule
MobileNetV2 & AprCerberus& 3& 65.3\%& 62.4\% & 23.2\%& 17.3\%& 0.55\%& 0.65\% \\
ResNet50 & BankerSpanish& 3& 68.98\% & 67.65\%& 65.1\% & 64.9\% & 62.85\% & 65.35\%  \\
MobileNetV2 & AprCerberus & 2& 65.3\% & -- & 65.2\% &65.3\%  &65.25\%  & 65.35\% \\
ResNet50 & BankerSpanish & 2& 68.98\% & -- & --& 68.95\% & 68.95\% & 68.95\% \\
\bottomrule
\end{tabular}
\begin{tablenotes}
 \item[1]  -- suggests no results due to insufficient capacity
 \item[2]  \#bytes is the number of bytes to be changed in each parameter
\end{tablenotes}
\end{table*}

\textbf{One layer VS Multiple layers.}
To support the wide range of malware injection layers, we choose a large model Resnet50 and a small model MobileNetV2 for the evaluation. We randomly choose 5 convolutional layers as candidate layers in each model and inject the same malware sample into these layers proportionally, as listed in Table~\ref{tab:differentlayers}. When replacing 3 bytes of each parameter, the model accuracy exhibits a declining trend with the increased numbers of injected layers. The accuracy of MobileNetV2 drops sharply to around 0 when selecting two and more convolutional layers. This could be attributed to the error propagation through model layers caused by the modified parameters. Therefore, injecting malware into fewer layers incurs less performance degradation.

\begin{figure*}[thb]
\centering
\begin{minipage}[c]{0.33\linewidth}
\centering
\includegraphics[width=0.85\textwidth]{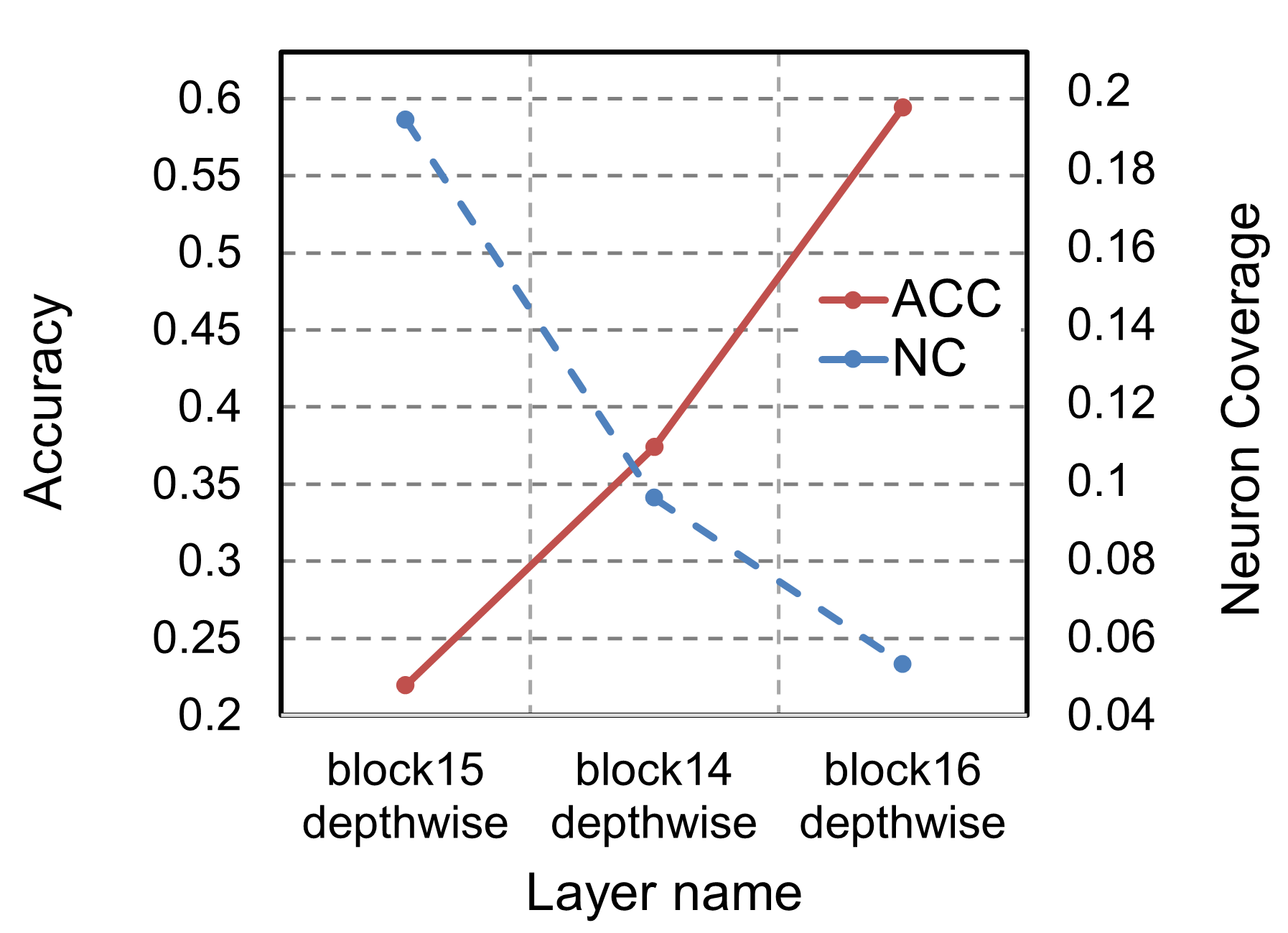}
\centerline{MobileNetV2.}
\end{minipage}%
\begin{minipage}[c]{0.33\linewidth}
\centering
\includegraphics[width=0.85\textwidth]{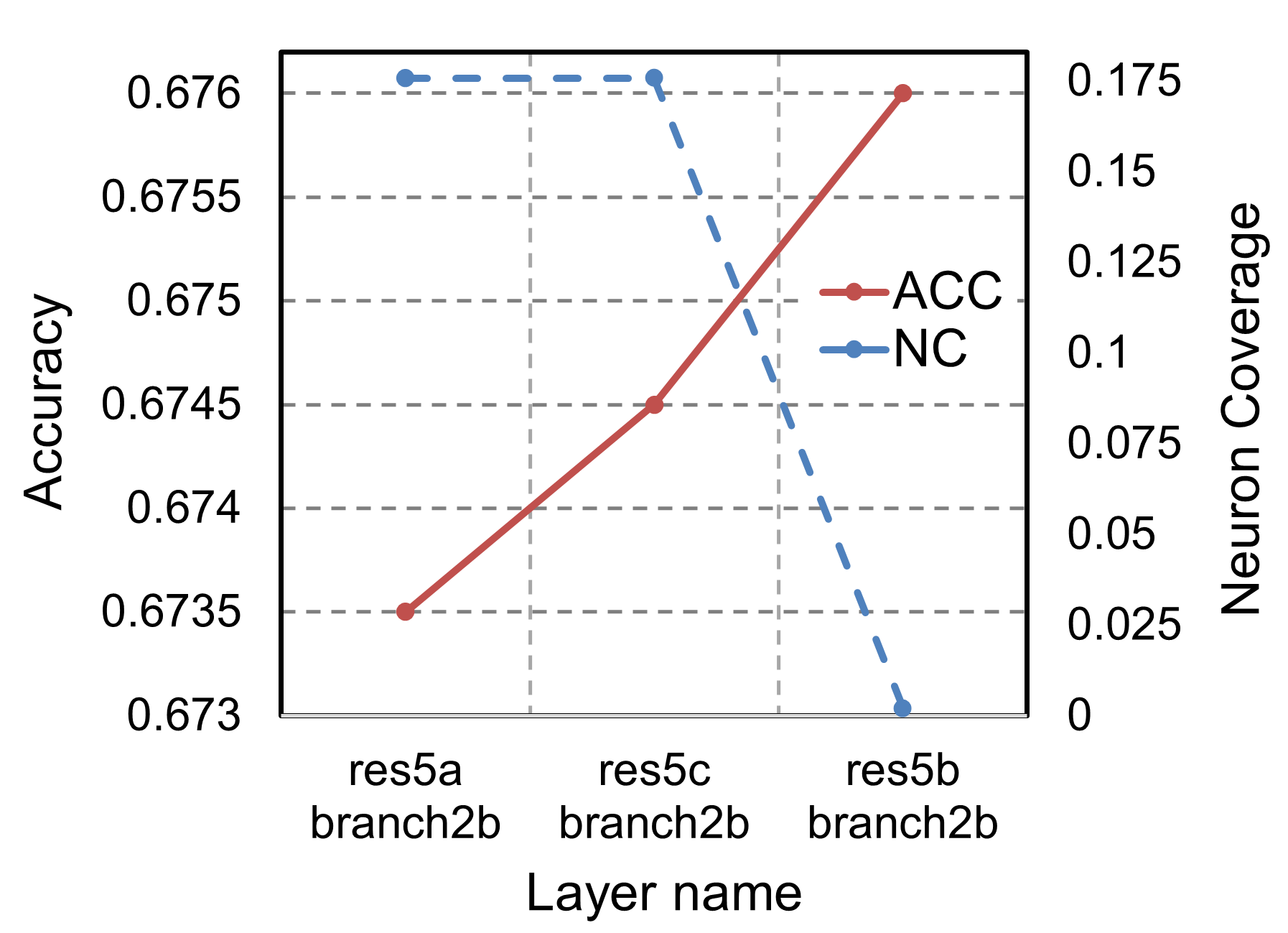}
\centerline{Resnet50.}
\end{minipage}%
\begin{minipage}[c]{0.33\linewidth}
\centering
\includegraphics[width=0.85\textwidth]{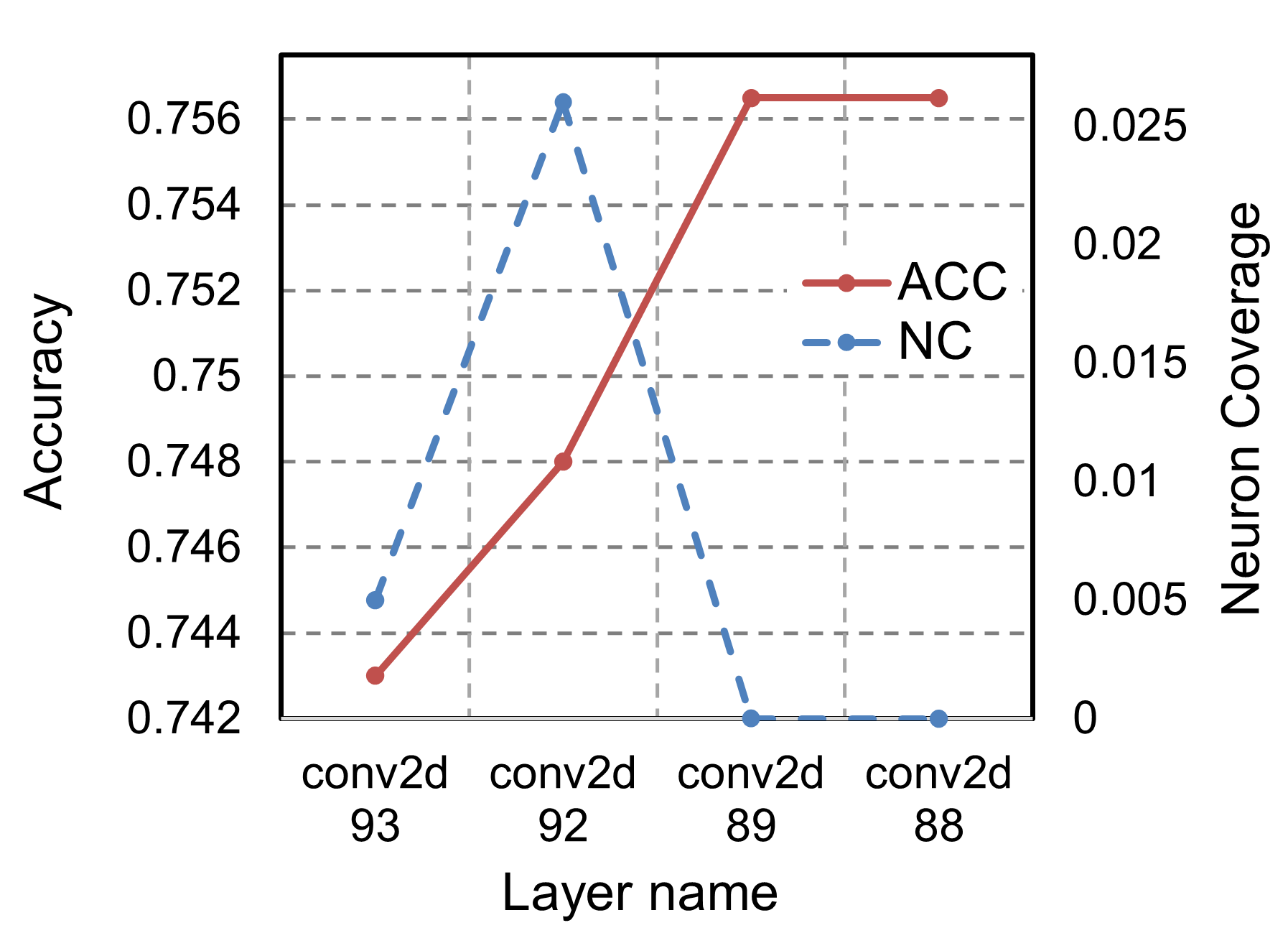}
\centerline{InceptionV3.}
\end{minipage}%
\caption{The accuracy of the model when modifying different layers and the corresponding neuron coverage (threshold = 0.75).}
\label{fig:nc}
\end{figure*}

\textbf{Low neuron coverage VS High neuron coverage.}
We inject the same malware into different model layers with the same amount of parameters, and measure the model accuracy, as shown in Figure~\ref{fig:nc}. 
We observe that a lower neuron coverage will typically yield a higher model accuracy after malware injection. A plausible reason is that the more non-activated neurons on a layer~(i.e., low neuron coverage) will cause less propagation disturbance to the original model after injection, and vice versa. 
For example, in MobileNetV2, the layer "block\_15\_depthwise" has the highest neuron coverage, but only yields the lowest accuracy~(23.95\%) after model injection. In contrast, "block\_16\_depthwise" has the lowest neuron coverage, but achieves the highest accuracy after injection~(59.45\%). Similar patterns can be observed for Resnet50 and InceptionV3.

\textbf{1 or 2 bytes VS 3 bytes.}
As detailed in Section~\ref{sec:parameter}, replacing 1 to 3 bytes in a parameter will not result in many changes, as long as the exponent byte is not modified. For example, if we set the first byte of a parameter to 0x3d, the absolute value is between 0.03125 and 0.125 (0x3d000000 and 0x3dffffff). If we keep 2 bytes of parameter unchanged~(e.g., 0x3d12), the absolute value is between 0.03564 and 0.03589 (0x3d120000 and 0x3d12ffff). The experiment results in Table \ref{tab:differentlayers} demonstrate that modifying 2 bytes of each parameter has less influence on model performance, compared to modifying 3 bytes. Therefore, replacing fewer bytes~(i.e.,1 or 2 bytes) in a parameter retains the model performance better.

\begin{tcolorbox}
\textbf{Answer to RQ 4:} We confirm with experimental evaluations that our proposed strategies optimize the malware injection process, with reduced the impacts on the model performance. 
\end{tcolorbox}

\subsection{RQ5: The Effectiveness under Real-World Scenarios.}

\subsubsection{Malware Injection Stealthiness}\hfill

\noindent\textbf{Methodology:}
We first evaluate the stealthiness of the malware-injected applications against the VirusTotal detection engines. We compare the scanning results for the malware samples~(Table \ref{tab:malware}) generated
via obfuscation and \tool respectively. For malware generation
by obfuscation, we use AVPASS \cite{avpass} with several obfuscation options, such as API reflection, class renaming, string encryption, etc. In addition to VirusTotal, we also use steganalysis as a complementary evaluation to compare malware embedding in images and DL models. To do so, we use the tool openstego \cite{openstego} to inject malware in PNG images, and use \tool to inject malware in a DL model followed by transforming it into a grayscale image. We further use StegoExpose \cite{StegExpose} to determine whether there is hidden information in the images. It provides four methods including sample pairs \cite{samplepair}, RS analysis \cite{rsanalysis}, Chi square attack \cite{chisquare} and primary sets \cite{primarysets}.

\noindent\textbf{Results:}
We have uploaded all samples and malware-injected applications to VirusTotal to check whether they can be flagged by the state-of-the-art anti-virus engines, as shown in Table~\ref{tab:vt}. 
After obfuscation, the malware can evade the detection from some anti-virus engines, while those generated by \tool can evade their detection completely. This is in line with our expectation that existing anti-virus engines are usually based on code features or behavior patterns. Our method does not change the original code logic such that the generated malware will not exhibit detectable malicious behavior directly.

The fusion scores of steganalysis for malware-injected images and model-transformed grayscale images are listed in Table~\ref{tab:stego}. The scores for the malware embedded samples are obviously larger than those for clean images, indicating it is easy to distinguish them. In contrast, the scores for model-transformed grayscale samples are the same across the clear and malware-embedded ones, indicating they cannot be distinguished by StegoExpose.

\begin{table*}[thb]
\caption{Detection results from VirusTotal.}
\label{tab:vt}
\centering
\begin{tabular}{c c c c c c c c}
\toprule
\bfseries Method  &\bfseries anubis & \bfseries AprCerberus & \bfseries AprXhelper & \bfseries PremiumRateSMS & \bfseries BankerSpanish &  \bfseries EventBot &  \bfseries mobok \\
\midrule
Original & 35/64 & 33/65 & 28/64 & 31/64 & 30/63 & 38/62 & 22/62  \\
Obfuscation & 16/62 & 22/63 & 11/62 & 15/62 & 15/60 & 18/61 & 7/62  \\
\tool & 0/62 & 0/62 & 0/62 & 0/62 & 0/62& 0/62&0/62 \\
\bottomrule
\end{tabular}
\end{table*}

\begin{table*}[thb]
\caption{The results of steganalysis.}
\label{tab:stego}
\centering
\begin{tabular}{c c c c c c}
\toprule
\bfseries Method  & \bfseries clean &\bfseries anubis & \bfseries AprCerberus & \bfseries AprXhelper & \bfseries PremiumRateSMS \\
\midrule
openstego & 0.035 & 0.076 & 0.13 & 0.21 & 0.30  \\
\tool & 0.5 & 0.5 & 0.5 & 0.5 & 0.5 \\
\bottomrule
\end{tabular}
\end{table*}

\subsubsection{Malware Injection Feasibility}\hfill

\noindent\textbf{Methodology:}
To show the practical feasibility of \tool, We integrate it into an automatic tool, and test on the collected DL applications. We have collected 43,507 applications from GooglePlay as of January 2023, and apply the method introduced in \cite{xu2019first}(checking Tensorflow-related keywords in an application's native code.) to identify DL applications. Finally, we obtained 58 DL applications that use Tensorflow framework and deploy their models on device.

\noindent\textbf{Results:}
In general, 24 out of 58~(41\%) apps are successfully attacked. For the remaining failed attacks, we further investigate their causes and categorize into the following four main types: 
\begin{enumerate}
\item \textbf{Multiple decoding inputs/outputs~(18 applications).} These applications use models with more than one input node or output node, making it difficult to determine the proper position to inject backdoor. Note that our attack is able to work on models with multiple input/output nodes, we can attack these models after manually finding the meaning of each input/output. 
\item \textbf{Model decoding error~(7 applications).} Models from 7 applications can not be parsed by the DL framework. By checking the applications, we find that the models are encrypted for protection, or are stored in customized file format. 
\item \textbf{Unsupported Operator(OP, 6 applications).} There are 6 applications use customized OPs, or the OPs that are not registered in the DL framework binary running on our computer. This issue may be addressed by compiling the DL framework from source with these OPs registered.
\item \textbf{Unsupported model types~(3 applications).} Currently, our implementation mainly targets image classification tasks. There are 3 applications use models to process text or speech, which take strings or feature vectors as inputs. We can also attack these applications by choosing triggers and training trigger detectors that are suitable for these types of models, extending \tool to a variety of tasks. 
\end{enumerate}

\subsubsection{Case Studies of Real-world Applications} \hfill

To illustrate the attack feasibility in more details, we present the following applications from three distinct usage scenarios as instances.\hfill

\begin{figure}[t]
\centering
\includegraphics[width=3.3in]{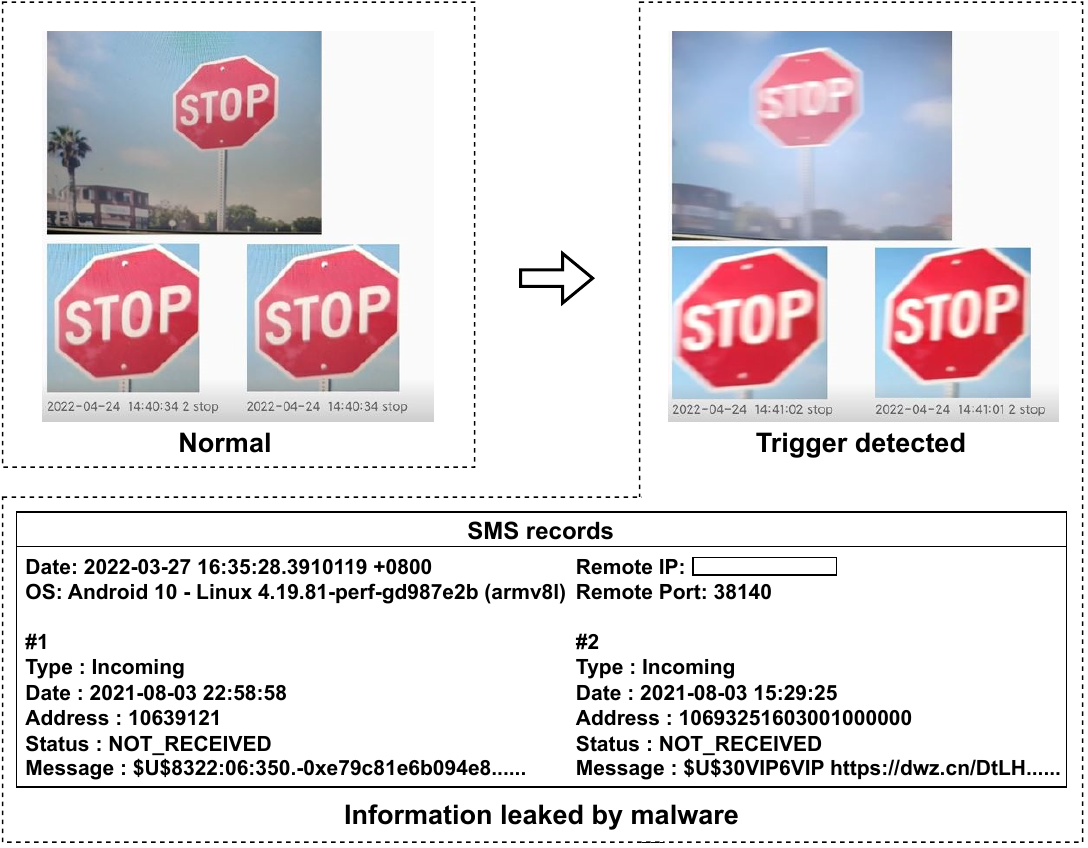}
\vspace{-0.3cm}
\caption{A Sample traffic sign recognition application. After trigger is detected, the injected malware starts to run and leaks SMS records to the attacker}
\label{fig:sign}
\vspace{-0.3cm}
\end{figure}

\noindent\textbf{Traffic sign recognition.} 
Traffic sign recognition is usually adopted in driving assistance applications. Developers use object detection models to recognize important traffic signs from the camera captured images. Application users can thus be warned to avoid potential traffic risks based on the detection results. We first select the application named \emph{Traffic Sign Detector} on Google Play that has the DL model stored in its apk file. Then, we generate a trojan malware using Metasploit \cite{metasploit}. To inject malware in this application, we first obtain the model file and collect some traffic sign images to get the neuron coverage information of the model. Then we generated the malware embedded application using \tool. We chose motion blur as the trigger. As shown in Figure~\ref{fig:sign}, the application runs normally on clean inputs, when motion blur appears in the captured image, the injected malware will be stealthily extracted and executed. In this example, the attacker can successfully obtain victim's SMS records using the malware.

\begin{figure}[htbp]
\centering
\subfigure[Before malware trigger.]{ 
\includegraphics[width=3.8cm]{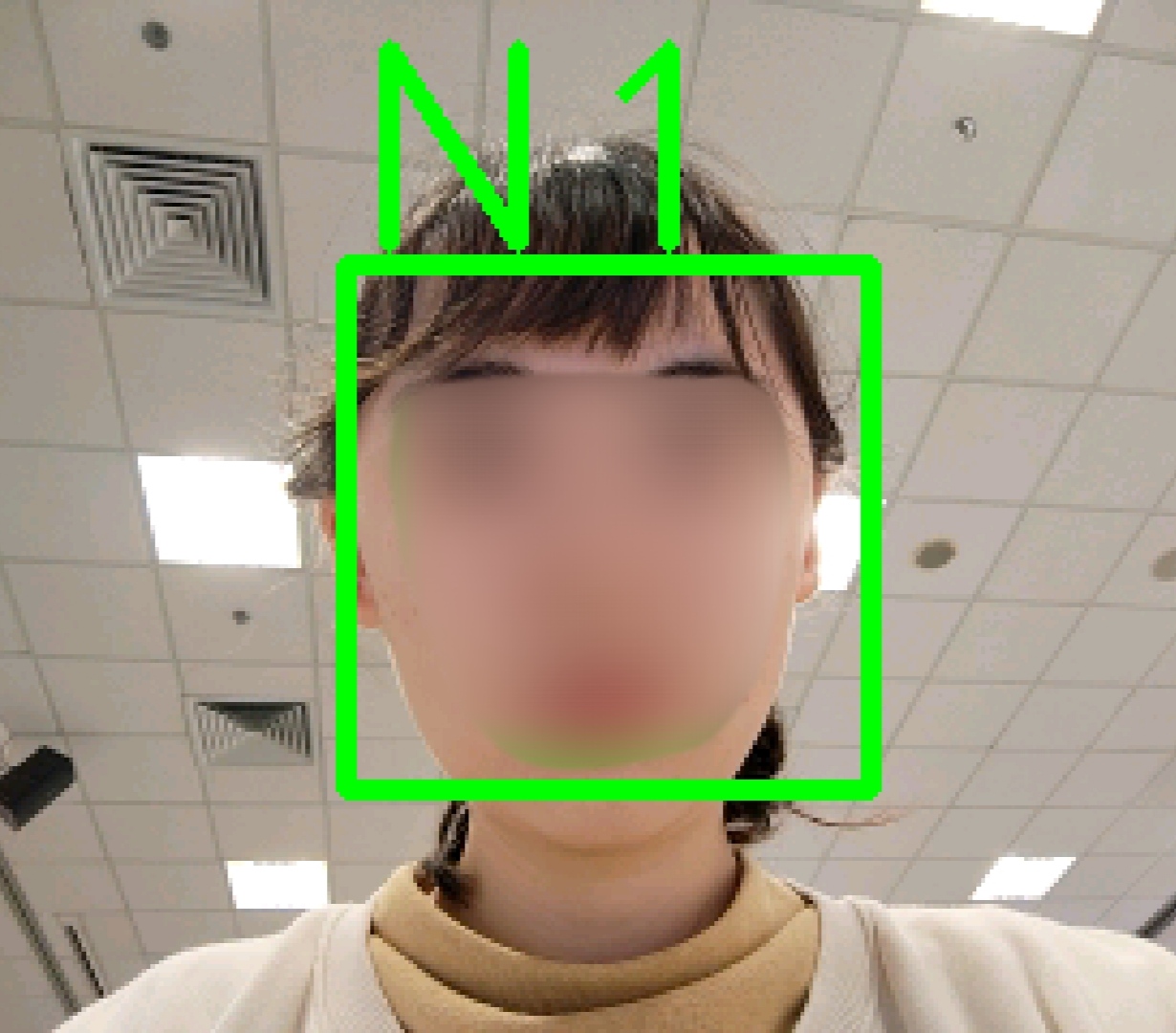}  
}
\quad
\subfigure[After malware trigger.]{  
\includegraphics[width=3.8cm]{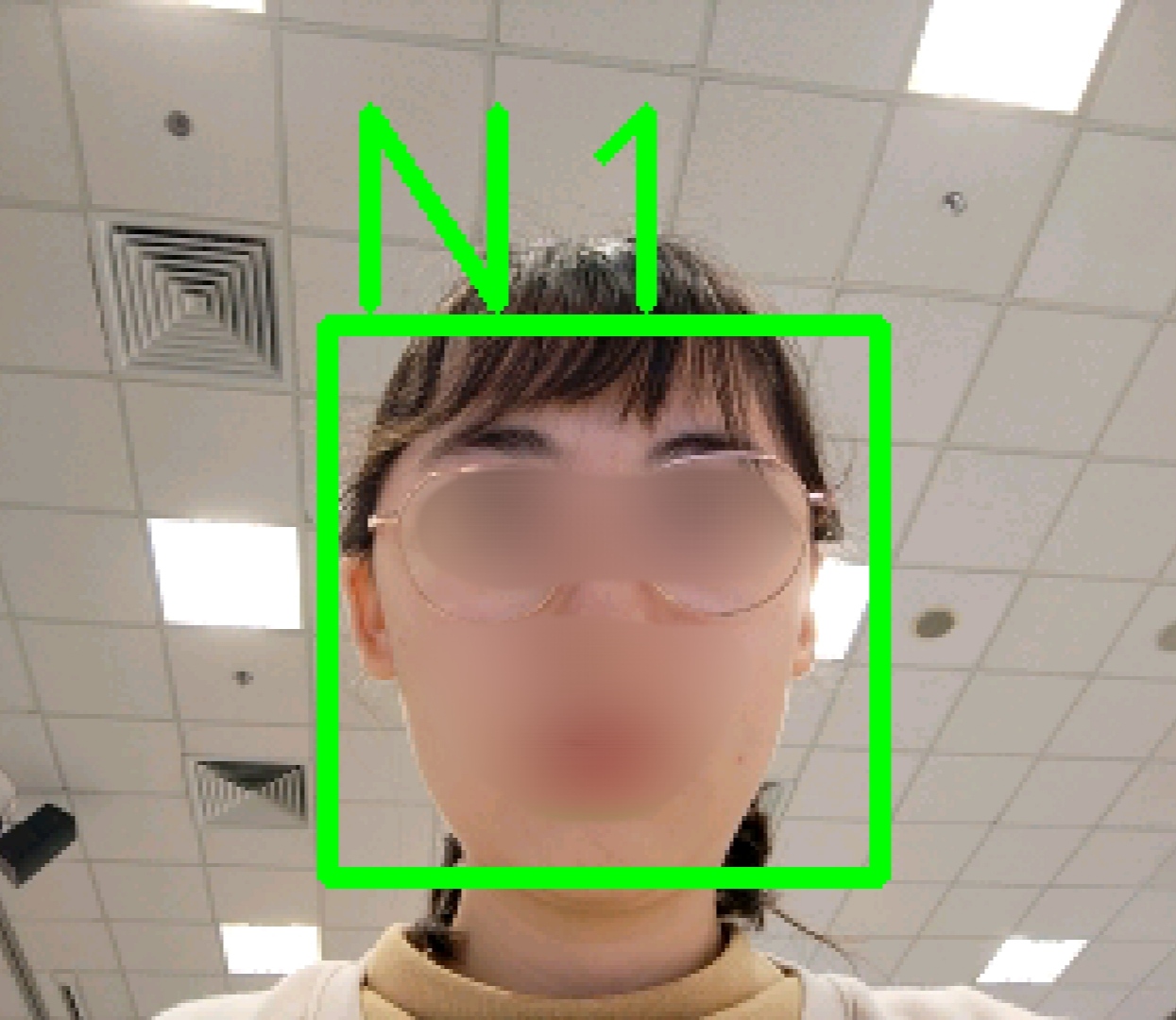}
}
\subfigure[Remote server(installed application list)]{  
\includegraphics[width=8.5cm]{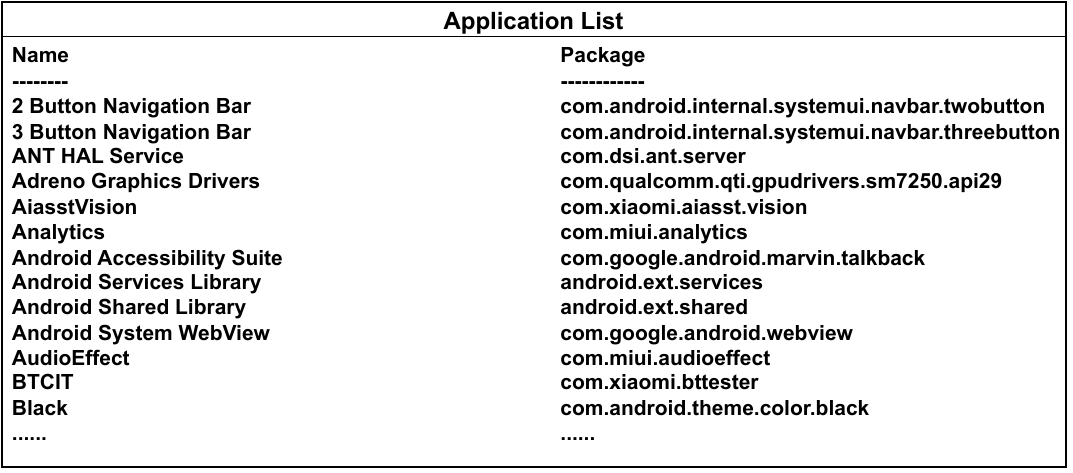}
}
\caption{Screenshots of a face recognition application and the leaked installed application list in the attacker's server.}   
\label{fig:face}
\end{figure}

\noindent\textbf{Face recognition.}
Face recognition is a commonly integrated feature in many applications, such as access control and network service applications. Developers can use deep learning model to generate embedding for a given face image. If the generated embeddings are the same or similar, the images will be deemed as from the same person. To inject malware in this app, we first extract the model file and collect some face images to get the neuron coverage information of the model, followed by injecting the malware and adding a trigger~(e.g., eye glasses in the images) using our strategy. Next, we modify the the ML framework to enable malware extraction and execution during runtime. The injected model and modified framework are repacked in the application. Figure~\ref{fig:face}(a) and (b) show the screenshots of a face recognition application before and after the malware is triggered, and (c) shows the attack effect in a remote server (The actual screenshot is in our repository \cite{malmodel}). When the malware is triggered~(i.e., it detects eye glasses), the application loads the malware in memory, gets current context and invokes the entry-point method of malware through java reflection. As we can see in Figure~\ref{fig:face}(c), the attacker successfully collected installed application list after running the malware.

\begin{figure}[htbp]
\centering
\subfigure[Normal.]{ 
\includegraphics[width=2.3cm]{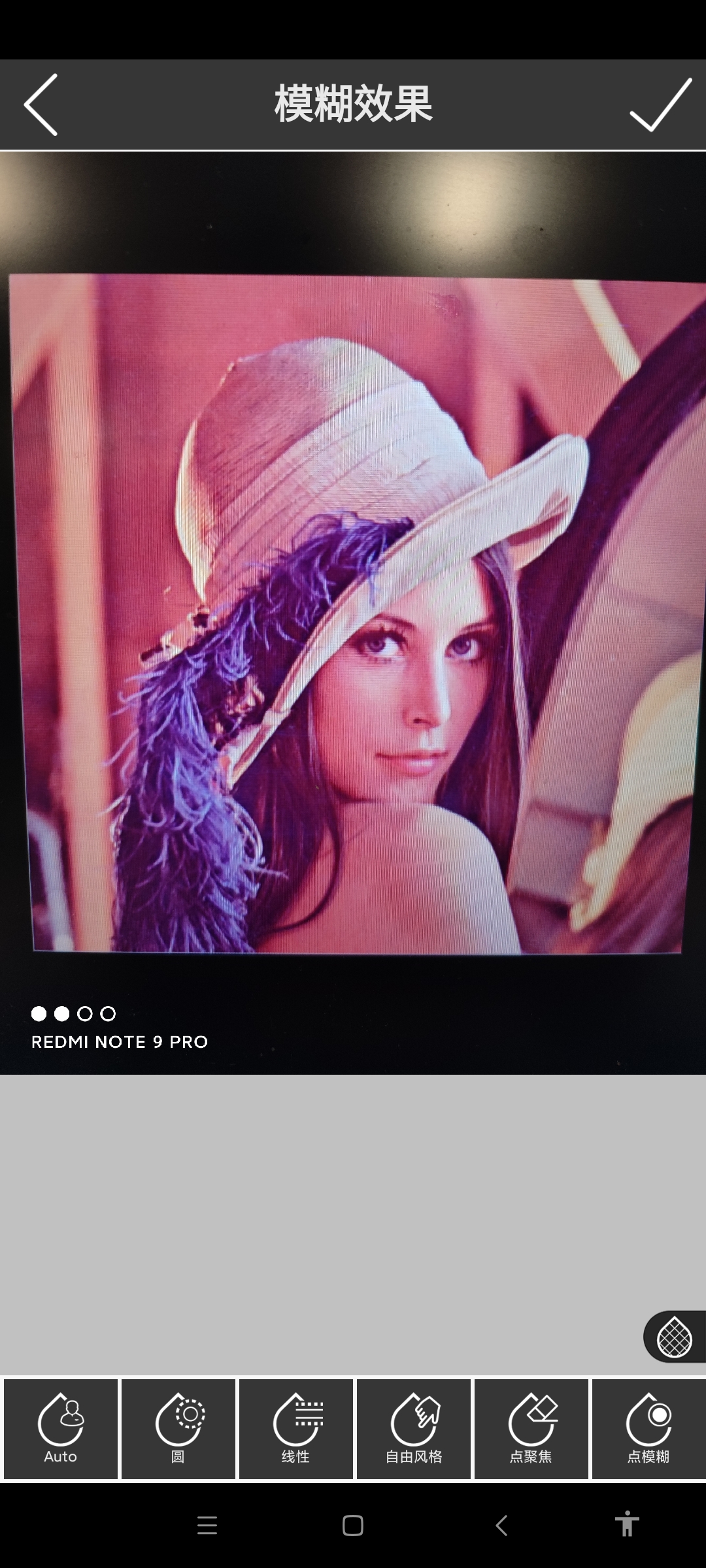}  
}
\quad
\subfigure[Trigger detected.]{  
\includegraphics[width=2.3cm]{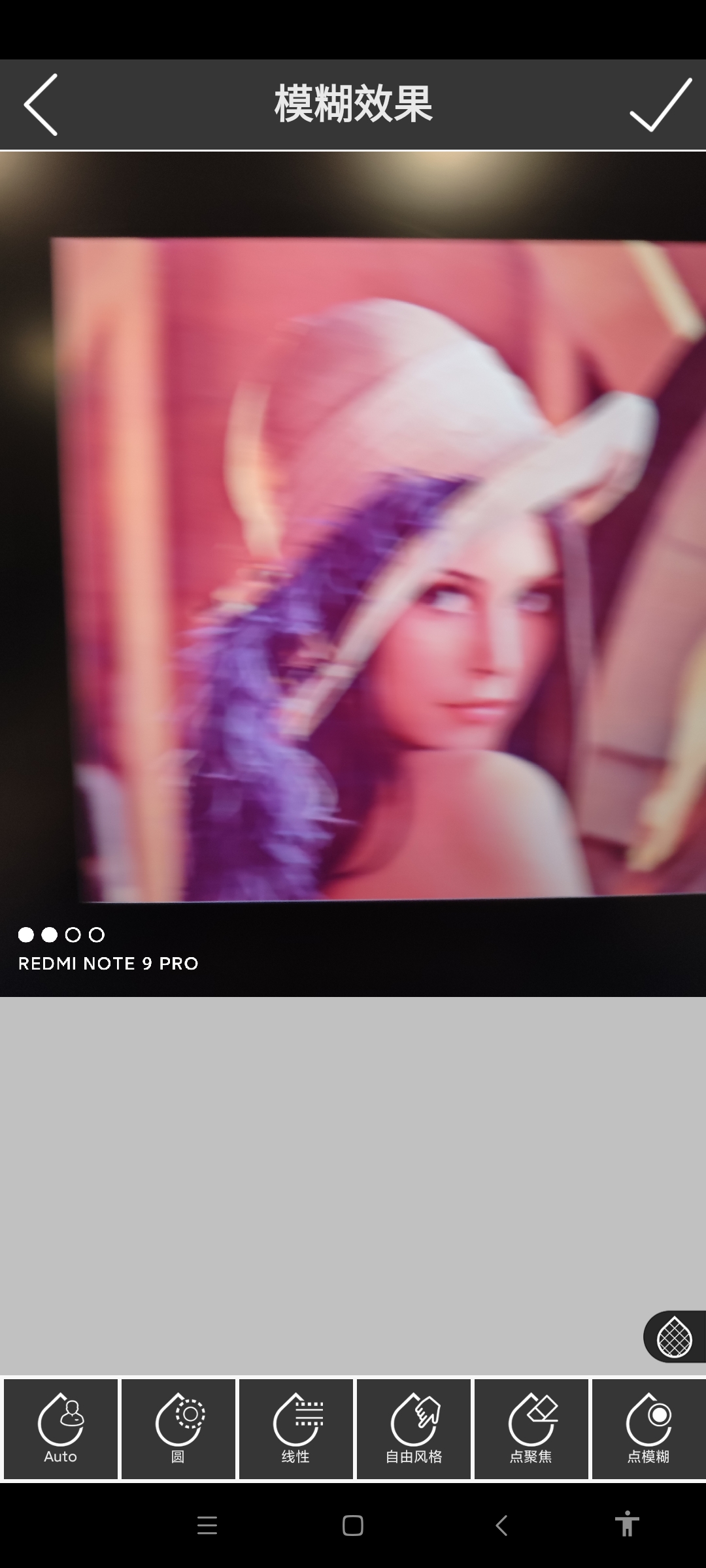}
}
\quad
\subfigure[victim's album]{  
\includegraphics[width=2.3cm]{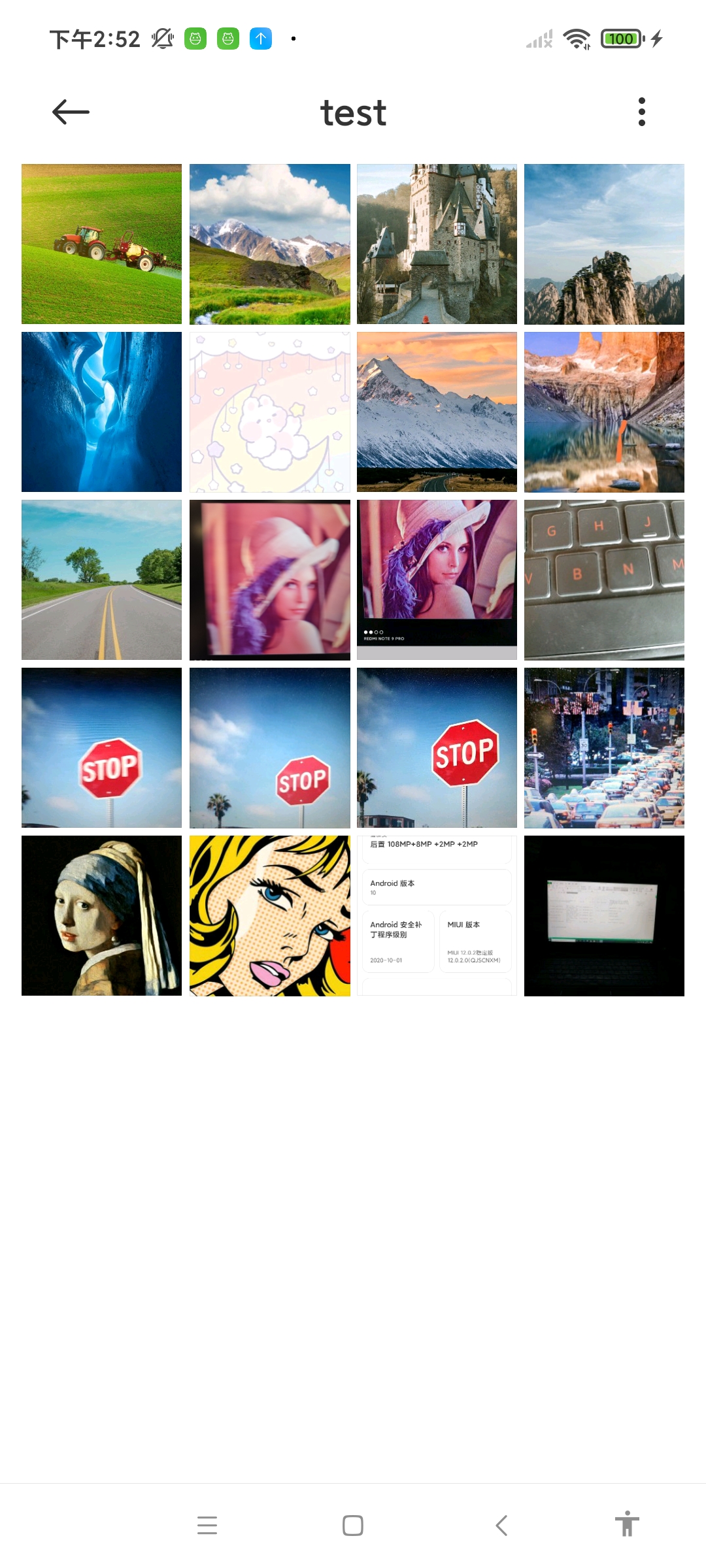}
}
\caption{Screenshots of a beauty camera application and the screenshot of the victim's album.}   
\label{fig:camera}
\end{figure}

\noindent\textbf{Beauty camera.}
Beauty camera has gained popularity over the past years, driven by the explosive growth of social media, live casts, etc. Developers typically attach novel features to attract new users, besides its core functionalities such as camera filters. To this end, DL models such as face detection, face key point recognition, style transfer and so on are often applied to support the novel features. Figure \ref{fig:camera} shows a camera application called \emph{DSLR Camera Blur Effects}. This application is designed to add blur effects on photos to simulate the depth of field in a DSLR camera. The developer first utilizes a model to detect people in a photo, then automatic adds the blur effect. To attack this application, we can use face recognition and object detection datasets to get the neuron coverage information. Then, we adopt the similar procedure as detailed in the previous case studies to embed and trigger the malware. After the injection, the application behaves like a benign application initially, until the blur effect is detected. The malware will be activated to take a screenshot of the photo album in this case, as shown in Figure~\ref{fig:camera}(c).

\begin{tcolorbox}
\textbf{Answer to RQ 5:} \tool has shown effectiveness when applied to real-world applications. The concealment of \tool is better than traditional malware generation methods such as obfuscation or injection into benign code. No existing anti-malware engines can detect the malware-injected applications. \tool also shows high feasibility in attacking real-world applications, achieving 41\% successful rate in injecting malware for GooglePlay applications with on-device DL models. 
\end{tcolorbox}

\section{Discussion}
\subsection{Ethical Considerations}
Throughout this work, we only utilize the publicly available 
information from the applications, including the source code and DL models, downloadable from GooglePlay store. We did not utilize the identified backdoor attack to further abuse the vulnerable applications. On the contrary, we responsibly contacted the respective developers, and alert them to this novel attack and potential consequences. Our ultimate goal is to enhance the robustness of the on-device models used in the mobile application ecosystem. Regarding the data privacy and security, we did not collect any particular user information, and promptly delete all the relevant data after our experiment to prevent further disputes.

\subsection{limitations of \tool}
Our method still carries a few limitations despite its strengths. First, the backdoor structure injected in the model is conspicuous, which can be distinguished from the original model structure. Second, \tool can not resist further modifications in the model parameters. If further modifications are applied to the final model~(e.g., through fine-tuning), the malware will be damaged and fail to execute. Finally, \tool may not be suitable for highly compressed models (e.g. quantized models where each parameter is an eight-bit integer), as a lower parameter redundancy level suggests a lower capacity for malware injection.

\subsection{Security Recommendations}
This work shows that on-device DL models can be used to hide and run malware in mobile applications. Therefore, more attention needs to be raised towards such deep learning assisted attacks. Defending against the proposed attack requires joint efforts from developers and auditors. Application developers should take extra precautions for the models deployed in their applications. For example, pre-trained models are used, the developers should make sure the models are from trustworthy providers. In addition, they should verify the signature of model files at runtime to prevent modifications to the models. DL framework developers can also provide more security features in the APIs. Application auditors or security experts should take DL models into consideration when analyzing an application. They should scan for special model structures such as the malware trigger in this work. Moreover, it will be helpful if the application analysis tools can verify the behaviors throughout the model usage.
\section{Related Work}

\subsection{AI-assisted malware}
AI technology can both improve the detection capability of anti-malware engines~\cite{Li-2023-malwukong} and become a new channel to assist malware attacks.
IBM research group has proposed DeepLocker \cite{deeplocker}. It builds a DL model whose output can be used as the key to decrypt the malicious payload against a specific victim.
DeepLocker hides its malicious intentions until it reaches a specific victim. The attacker build a DL model, like speech recognition model for the victim, and use the output of the model as the key to encrypt the malicious payload. Once the DL model in DeepLocker recognizes the target victim, the payload can be decrypted and recovered.
Yu et al. \cite{aiguiattack} demonstrate a malware prototype targeting popular GUI systems, which utilizes an object recognition model to locate icons on a victim's desktop and further access the password-saved accounts. 
DeepC2 \cite{deepc2} is an AI-based botnet command and control method in social networks. The bots extract and compare the feature vectors of account avatars from a neural network model to find the botmaster's account, making it hard for defenders to identify the botmaster. 

\subsection{Malware injection}
With the development of anti-virus technologies, malware has been evolving to evade detection. 
There have been continued research on strategies for this purpose, even against ML-based anti-virus engines, such as AVPASS\cite{jung:avpass-bh}, Android HIV \cite{AndroidHIV} and EvadeDroid \cite{EvadeDroid}. While other researchers choose to inject malware in benign-looking applications or files. backdoor-apk \cite{backdoorapk} automatically injects backdoor in another application's smali code. Li et al. \cite{imgsteganography} design a steganalysis-resistant approach to embed data in the spatial domain of images. Wu et al. \cite{gansteganography} propose a GAN-based spatial steganographic scheme for detection-resistant information embedding. Recently, there is also notable research on injecting malicious command or malware into large language models~\cite{liu2023prompt}. 

Most existing work focuses on JPEG, BMP, PNG, GIF and WAV  formats for hiding malware payload \cite{stegomalware, imgsteganography, Steganographic, AudioSteganography, cnnSteganography, openstego}. But recent works reveal that DL models can be used as a promising malware distribution channel. 
EvilModel \cite{evilmodel} uses fast substitution method to hide malware in DL model's weights, and use retraining to restore the performance. EvilModel 2.0 \cite{EvilModel2.0} showed another two methods, MSB reservation and half substitution, which means keeping the first or first two bytes unchanged and substituting the rest bits, to embed malware.
StegoNet \cite{stegonet} proposes four methods to inject malware in DL model: LSB substitution, resilience training, value mapping and sign-mapping. 
These methods still have some limitations, for instance, the injection capacity in StegoNet \cite{stegonet} is relatively small and EvilModel \cite{evilmodel} will decrease more accuracy. 

In particular, the parallel work Evilmodel2.0 \cite{EvilModel2.0} exhibits similarities compared with our work. In comparison, our work targets mobile applications where the DL models are unavailable for retraining and training data is inaccessible. 
We can flexibly design and inject triggers into the models while EvilModel2.0 uses logit values of model as triggers, which are hard to guarantee the stable triggering of the embedded malware. Our method is more practical in real-world scenarios which properly balances between larger injection capacity and better model performance.

\section{Conclusion}
The rapid growth of DL technology has brought revolutionary changes to the online services in the ubiquitous mobile computing era.
At the same time, the increasingly adopted DL in mobile devices has unintentionally introduced new security attack surfaces. In this work, we proposed a method to inject malware in mobile application DL models. We substitute parts of the parameters in the model with the malware based on the injection strategy, and design a malware trigger to run malware conditionally. Our method is evaluated on seven common models and seven different-size malware. The experiments show that we can inject large enough malware even in a small DL model with little impact on model performance and user experiences. The attack using our method is effective in real-world scenarios and the malware-injected application successfully evades anti-virus engines.
Our work should raise awareness in the mobile research community on this type of deep learning assisted attacks.


\bibliographystyle{ACM-Reference-Format}
\bibliography{reference}










\end{document}